\documentclass[12pt]{article}
\usepackage[dvips]{graphicx}
\usepackage[dvipdfm,colorlinks=true,linkcolor=blue,citecolor=blue]{hyperref}
\usepackage[round]{natbib}
\usepackage{amsmath,amsthm,amssymb,bm,hypernat} 
\usepackage{setspace}
\usepackage[ruled]{algorithm2e}
\newcommand{\argmin}{\mathop{\rm argmin}\limits}
\usepackage[letterpaper, margin=1in]{geometry}

\title{Robust and sparse Gaussian graphical modeling under cell-wise contamination}
\author{Shota Katayama$^1$, Hironori Fujisawa$^{2,3}$ and Mathias Drton$^4$}
\date{\vspace{-5ex}}

\begin{document}
\maketitle
\footnotesize 
\begin{center}
$^1$Department of Industrial Engineering and Economics, Tokyo Institute of Technology, Japan \\ 
$^2$The Institute of Statistical Mathematics, Japan \\
$^3$Nagoya University Graduate School of Medicine, Japan \\
$^4$Department of Statistics, University of Washington, USA

\end{center}
\normalsize
\begin{abstract}
  Graphical modeling explores dependences among a collection of
  variables by inferring a graph that encodes pairwise conditional
  independences.  For jointly Gaussian variables, this translates into
  detecting the support of the precision matrix.  Many modern
  applications feature high-dimensional and contaminated data that
  complicate this task.  In particular, traditional
  robust methods that down-weight entire observation vectors are often
  inappropriate as high-dimensional data may feature partial
  contamination in many observations.  We tackle this problem by
  giving a robust method for sparse precision matrix estimation based
  on the $\gamma$-divergence under a cell-wise contamination model.
  Simulation studies demonstrate that our procedure
  outperforms existing methods especially for highly contaminated
  data.
\end{abstract}

\textbf{keywords}:
cell-wise contamination; Gaussian graphical modeling;
  precision matrix; sparsity; robust inference

\section{Introduction}\label{sec1}

Let $\bm{Y}=(Y_{1},\dots ,Y_{p})^{T}$ be a $p$-dimensional random
vector representing a multivariate observation.  The conditional
independence graph of $\bm{Y}$ is the undirected graph $G=(V,E)$ whose
vertex set $V=\{1,\dots, p\}$ indexes the individual variables and
whose edge set $E$ indicates conditional dependences among them.  More
precisely, $(i,j)\not\in E$ if and only if $Y_{i}$ and $Y_{j}$ are
conditionally independent given
$Y_{V\backslash\{i,j\}}=\{Y_k:k\not= i,j\}$.  For a Gaussian vector, the
edge set $E$ corresponds to the support of the precision matrix.
Indeed, it is well known that if $\bm{Y}$ follows a multivariate
Gaussian distribution $N_{p}(\bm{\mu},\bm{\Sigma})$ with mean vector
$\bm{\mu}$ and covariance matrix $\bm{\Sigma}$, then $(i,j)\not\in E$
if and only if $\bm{\Omega}_{ij} = 0$, where
$\bm{\Omega} = \bm{\Sigma}^{-1}$.

Inference of the conditional independence graph sheds light on direct
as opposed to indirect interactions and has received much recent
attention \citep{drton17}.  In particular, for high-dimensional
Gaussian problems, several techniques have been developed that exploit
available sparsity in inference of the support of the precision matrix
$\bm{\Omega}$.  \citet{meinshausen06} suggested fitting node-wise
linear regression models with $\ell_{1}$ penalty to recover the
support of each row.  \citet{yuan07},
\citet{banerjee08} and \citet{friedman08} considered the graphical
lasso (Glasso) that involves the $\ell_{1}$ penalized log-likelihood
function.  \citet{cai11} proposed the constrained $\ell_{1}$
minimization for inverse matrix estimation (CLIME), which may be formulated
as a linear program.  
Yet
other approaches can be found in \citet{yuan09}, \citet{peng09},
\citet{zhang14}, \citet{khare15}, \citet{liu15}, and \citet{lin16}.

In fields such as bioinformatics and economics, data are often not
only high-dimensional but also subject to contamination.  While
suitable for high dimensionality, the above mentioned
techniques are sensitive to contamination.  Moreover, traditional
robust methods may not be appropriate when the number of variables is
large.  Indeed, they are based on the model in which an observation
vector is either without contamination or fully contaminated.  Hence,
an observation vector is treated as an outlier even if only one of
many variables is contaminated.  As a result these methods down-weight
the entire vector regardless of whether it contains `clean' values for
some variables.  Such information loss can become fatal as the
dimension increases.  As a more realistic model in high dimensional
data, \citet{alqallaf02} considered cell-wise contamination: The
observations $\bm{X}_{1},\dots, \bm{X}_{n}$ with p variables are generated by
\begin{align}\label{cmodel}
  \bm{X}_{i} = (\bm{I}_{p} - \bm{E}_{i})\bm{Y}_{i} + \bm{E}_{i}\bm{Z}_{i},\quad i=1,\dots, n.
\end{align}
Here, $\bm{I}_{p}$ is the $p\times p$ identity matrix and each
$\bm{E}_{i} = {\rm diag}(E_{i1},\dots, E_{ip})$ is a diagonal random
matrix with the $E_{ij}$'s independent and Bernoulli distributed with
$P(E_{ij} = 1) = \varepsilon_{j}$.  The random vectors $\bm{Y}_{i}$
and $\bm{Z}_{i}$ are independent, and
$\bm{Y}_{i}\sim N_{p}(\bm{\mu}, \bm{\Sigma})$ corresponds to a clean
sample while $\bm{Z}_{i}$ makes contaminations in some elements of
$\bm{X}_{i}$.

Our goal is to develop a robust estimation method for the conditional
independence graph $G$ of $\bm{Y}_{i}$ from the cell-wise contaminated observations $\bm{X}_{i}$.  
Techniques such as node-wise regression, Glasso and CLIME process an estimate of the
covariance matrix.  Our strategy is thus simply to apply these
procedures using a covariance matrix estimator that is robust against
cell-wise contamination.  However, while many researchers have
considered the traditional `whole-vector' contamination framework
\citep[see, e.g.,][Chapter 6]{maronna06}, there are fewer existing
methods for cell-wise contamination.  Specifically, we are aware of
three approaches, namely, use of alternative $t$-distributions
\citep{fine11,fine14}, use of rank correlations
\citep{ollerer15,loh15}, and a pairwise covariance estimation method
by \citet{tarr16} who adopt an idea of \citet{gnanadesikan72}.  In
contrast, in this paper, we provide a robust covariance matrix
estimator via $\gamma$-divergence as proposed by \citet{fujisawa08}.
The $\gamma$-divergence can automatically reduce the impact of
contaminations, and it is known to be robust even when the number of
contaminations is large.

The rest of this paper is structured as follows.  We review some graph
estimation methods in Section \ref{sec2.1} and the $\gamma$-divergence
in Section \ref{sec2.2}. In Section \ref{sec3}, the robust covariance
matrix estimator via $\gamma$-divergence is proposed and some of the
existing competitors are introduced.  Numerical experiments that
illustrate the benefits of our new method are presented in Section
\ref{sec4}.  Concluding remarks are given in Section \ref{sec5}.

\section{Preliminaries}\label{sec2}
\subsection{Graph estimation}\label{sec2.1}

For concise presentation, we focus on node-wise regression, Glasso and
CLIME.  Let $\hat{\bm{\Sigma}}$ be an estimator of $\bm{\Sigma}$.
For index sets $A$ and $B$, define $\hat{\bm{\Sigma}}_{A,B}$ as the
sub-matrix of $\hat{\bm{\Sigma}}$ with the rows in $A$ and the columns
in $B$.  We use the shorthand $\setminus j$ for the set
$V\backslash \{j\}$, so that
$\hat{\bm{\Sigma}}_{\backslash j, \backslash j}$ denotes the
sub-matrix with both rows and columns in $V\backslash \{j\}$.  In
$\ell_{1}$ penalized node-wise regression, one finds
\begin{align*}
\hat{\bm{\beta}}^{(j)} = \argmin_{\bm{\beta}\in \mathbb{R}^{p-1}}
\frac{1}{2}\bm{\beta}^{T}\hat{\bm{\Sigma}}_{\backslash j, \backslash j}\bm{\beta}
-\hat{\bm{\Sigma}}_{j, \backslash j}\bm{\beta} + \lambda \|\bm{\beta}\|_{1},\quad
j = 1,\dots, p,
\end{align*}
where the tuning parameter $\lambda > 0$ controls the strength of the
penalty $\|\bm{\beta}\|_{1}$.  Large $\lambda$ yields high sparsity of
$\hat{\bm{\beta}}^{(j)}$.  After obtaining
$\hat{\bm{\beta}}^{(1)},\dots, \hat{\bm{\beta}}^{(p)}$, the edge set
$E$ is estimated by the ``AND'' rule
$\hat{E} = \{(i,j): \hat{\bm{\beta}}^{(j)}_{i} \neq 0 {\rm\ and\ }
\hat{\bm{\beta}}^{(i)}_{j}\neq 0\}$ or the ``OR'' rule
$\hat{E} = \{(i,j): \hat{\bm{\beta}}^{(j)}_{i} \neq 0 {\rm\ or\ }
\hat{\bm{\beta}}^{(i)}_{j}\neq 0\}$.  Node-wise regression is
well-defined for any positive semidefinite estimate $\hat{\bm{\Sigma}}$.

The Glasso estimator is obtained by solving
\begin{align}
  \label{eq:glasso}
\hat{\bm{\Omega}}^{{\rm Glasso}} = \argmin_{\bm{\Omega} \in \mathbb{R}^{p\times p}} {\rm tr}\bm{\hat{\Sigma}}\bm{\Omega}
 - \log |\bm{\Omega}| + \lambda \|\bm{\Omega}\|_{1},
\end{align}
where $\|\bm{\Omega}\|_{1}$ is the element-wise $\ell_{1}$ norm of
$\bm{\Omega}$ and $\lambda > 0$ is a tuning parameter that controls
the sparsity of $\bm{\Omega}$.  The edge set may be estimated by
$\hat{E}=\{(i,j): \hat{\bm{\Omega}}_{ij} \neq 0\}$.  Efficient
algorithms for the Glasso are given in \citet{friedman08} and
\citet{hsieh11}.  For convergence, the former requires
$\hat{\bm{\Sigma}} + \lambda \bm{I}_{p}$ to be positive semidefinite
while the latter requires the same for $\hat{\bm{\Sigma}}$.

Finally, we review the CLIME method. Let 
\begin{align}
  \label{eq:clime}
\hat{\bm{\Omega}}^{0}= \argmin_{\bm{\Omega}\in\mathbb{R}^{p\times p}} \|\bm{\Omega}\|_{1}
\quad {\rm subject\ to}\quad  \|\hat{\bm{\Sigma}} \bm{\Omega} - \bm{I}_{p}\|_{\infty} \le \lambda,
\end{align}
where $\|\cdot\|_{\infty}$ means the element-wise infinity norm.
Generally, $\hat{\bm{\Omega}}^{0} = (\hat{\omega}_{ij}^{0})$ is not
symmetric.  The CLIME is defined through a simple symmetrization, namely,
\begin{align*}
\hat{\bm{\Omega}}^{{\rm CLIME}}_{ij}
= \hat{\omega}_{ij}^{0}I(|\hat{\omega}_{ij}^{0}| \le |\hat{\omega}_{ji}^{0}|)
+\hat{\omega}_{ji}^{0} I(|\hat{\omega}_{ij}^{0}| > |\hat{\omega}_{ji}^{0}|),\quad
i,j = 1,\dots, p,
\end{align*}
and the edge set is estimated as in Glasso.  \citet{cai11} translated
the matrix optimization problem from~\eqref{eq:clime} into $p$ vector
optimization problems.  Each of them can be solved by linear
programming.  The CLIME essentially needs the positive definiteness of
$\hat{\bm{\Sigma}}$.  Without it, the optimization problem may be
infeasible or return inadequate solutions.

\subsection{Robust inference via $\gamma$-divergence}\label{sec2.2}
Let $f_{}$ and $f_{n}$ be the data generating and empirical densities,
respectively.  In robust inference one typically assumes that
$f = (1-\varepsilon) g + \varepsilon h$, where $g$ is the density of
clean data, $h$ is the density of contamination, and
$\varepsilon \ge 0$ is the contamination level.  For estimation of
$g$, consider a model with densities $g_{\theta}$ indexed by the
parameter $\theta$.  The Kullback-Leibler (KL) divergence between
$f_{n}$ and $g_{\theta}$ results in a biased estimate unless
$\varepsilon = 0$. To overcome it, \citet{fujisawa08} proposed the
$\gamma$-divergence given by
\begin{align*}
d_{\gamma}(f_{n}, g_{\theta}) = -\frac{1}{\gamma} \log \int f_{n}(x) g_{\theta}(x)^{\gamma} dx
+\frac{1}{1+\gamma}\log \int g_{\theta}(x)^{1+\gamma}dx,
\end{align*}
where $\gamma > 0$ is a constant that controls the trade-off between
efficiency and robustness.  In fact, the $\gamma$-divergence is
equivalent to the KL divergence when $\gamma \to 0$.  The estimator
given by minimizing $d_{\gamma}(f_{n}, g_{\theta})$ over a possible
parameter space is highly robust.  Roughly speaking, in the limiting
case $n \to \infty$, $d_{\gamma}(f_{n}, g_{\theta})$ can be regarded
as $d_{\gamma}(f, g_{\theta})$, and 
\begin{align*}
d_{\gamma}(f, g_{\theta}) = -\frac{1}{\gamma}\log\bigg\{(1-\varepsilon)\int g(x) g_{\theta}(x)^{\gamma}dx + \varepsilon \nu(\theta; \gamma)\bigg\} + \frac{1}{1+\gamma}\log \int g_{\theta}(x)^{1+\gamma}dx,
\end{align*}
where $\nu(\theta; \gamma) = \int h(x)g_{\theta}(x)^{\gamma}dx$.  The
$\gamma$-divergence successfully provides robust estimates whenever
$\nu(\theta; \gamma) \approx 0$ over the parameter space
considered. In such a case, we see that
$d_{\gamma}(f,g_{\theta})\approx d_{\gamma}(g,g_{\theta}) -
(1/\gamma)\log(1-\varepsilon)$, where the contamination density $h$ is
automatically ignored, so that the minimizer of
$d_{\gamma}(f,g_{\theta})$ is approximately equal to the minimizer of
$d_{\gamma}(g, g_{\theta})$.  This is a favorable property, because
when $g=g_{\theta^{*}}$, the minimizer of $d_{\gamma}(g,g_{\theta})$
is $\theta^{*}$.  Fortunately, $\nu(\theta; \gamma)$ is close to zero
when $h$ lies in the tail of $g_{\theta}^{\gamma}$.  To illustrate
this fact, assume for a moment that $g_{\theta}$ is the density of
$N(\theta, 1)$.  If $h$ is the density of $N(\alpha, 1)$, then
$\nu(\theta;
\gamma)=c_{1,\gamma}\exp\{-c_{2,\gamma}(\alpha-\theta)^{2}\}$ for some
$c_{1,\gamma},c_{2,\gamma} > 0$.  Thus,
$\nu(\theta; \gamma)$ is small whenever $\alpha$ is not too close to
the set of parameters $\theta$ that determine the better fitting
densities $g_\theta$.

\section{Methods}\label{sec3}
\subsection{Proposed methodology}

As noted in Section \ref{sec2.1}, we seek a robust covariance estimate
$\hat{\bm{\Sigma}}$ for use in graph estimation.  In this section, we
construct such an estimate via the $\gamma$-divergence.  Our estimator
$\hat{\bm{\Sigma}}$ is constructed in an element-wise fashion and
exhibits robustness to cell-wise contamination.  We begin by writing
each covariance as
\begin{align}\label{deco}
\bm{\Sigma}_{jk} = \sqrt{\sigma_{jj}}\sqrt{\sigma_{kk}}\rho_{jk},
\end{align}
where $\sigma_{jj} = {\rm Var}(Y_{ij})$, $\sigma_{kk}={\rm Var}(Y_{ik})$ and $\rho_{jk} = {\rm Corr}(Y_{ij}, Y_{ik})$,
for $j,k = 1,\dots, p$. Here, $\bm{Y}_{i}=(Y_{i1},\dots, Y_{ip})^{T}$
is the $i$-th unobserved clean sample in (\ref{cmodel}).
We now derive estimates of the variances and the correlation
in~(\ref{deco}) based on the observations $\bm{X}_{i}=(X_{i1},\dots,
X_{ip})^{T}$, which under cell-wise contamination may have some of
their elements corrupted.

Fixing a coordinate $j\in\{1,\dots,p\}$, let
$g_{(\mu_{j}, \sigma_{jj})}$ be the density of
$N(\mu_{j}, \sigma_{jj})$, and let $f_{n}^{(j)}$ be the empirical
density of $X_{1j},\dots, X_{nj}$.  The robust estimators of $\mu_{j}$
and $\sigma_{jj}$ based on $\gamma$-divergence are given by
\begin{gather*}
(\hat{\mu}_{j}, \hat{\sigma}_{jj}) = \argmin_{\mu_{j}, \sigma_{jj}} d_{\gamma}(f_{n}^{(j)}, g_{(\mu_{j}, \sigma_{jj})}), \\
d_{\gamma}(f_{n}^{(j)}, g_{(\mu_{j}, \sigma_{jj})}) = -\frac{1}{\gamma}\log \sum_{i=1}^{n}\exp\bigg\{-\frac{\gamma}{2\sigma_{jj}}(X_{ij} - \mu_{j})^{2}\bigg\} + \frac{1}{2(1+\gamma)}\log \sigma_{jj}.
\end{gather*}
\citet{fujisawa08} gave an efficient iterative algorithm to compute
$(\hat{\mu}_{j}, \hat{\sigma}_{jj})$.  Let $\mu_{j}^{t}$ and
$\sigma_{jj}^{t}$ denote the $t$-th values starting from
initializations $\mu_{j}^{0}$ and $\sigma_{jj}^{0}$.  The algorithm
repeats the following steps until convergence:
\begin{align*}
\mu_{j}^{t+1} \leftarrow \sum_{i=1}^{n}w_{ij}^{t}X_{ij},\quad
\sigma_{jj}^{t+1} \leftarrow (1+\gamma)\sum_{i=1}^{n}w_{ij}^{t}(X_{ij} - \mu_{j}^{t+1})^{2},
\end{align*}
where the weights are updated as
\begin{align*}
w_{ij}^{t} = \exp\bigg\{-\frac{\gamma}{2\sigma_{jj}^{t}}(X_{ij} - \mu_{j}^{t})^{2}\bigg\}\bigg/\sum_{i=1}^{n}\exp\bigg\{-\frac{\gamma}{2\sigma_{jj}^{t}}(X_{ij} - \mu_{j}^{t})^{2}\bigg\}.
\end{align*}
We take the median of $X_{1j},\dots, X_{nj}$ as $\mu_{j}^{0}$ and the
median absolute deviation (MAD) as $\sigma_{jj}^{0}$.  

After $\hat{\mu}_{j}$ and $\hat{\sigma}_{jj}$ are obtained for
$j = 1,\dots, p$, we estimate each correlation $\rho_{jk}$ from the
standardized observations
$Z_{ij} = (X_{ij} - \hat{\mu}_{j})/\sqrt{\hat{\sigma}_{jj}}$.  Let
$h_{\rho_{jk}}$ be the bivariate standardized normal density
with correlation $\rho_{jk}$, and let $f_{n}^{(j,k)}$ be the empirical
density of $(Z_{1j}, Z_{1k}), \dots, (Z_{nj}, Z_{nk})$.  Our
correlation estimator is
\begin{gather}
\label{eq:gamma-rho}
\hat{\rho}_{jk} = \argmin_{|\rho_{jk}| < 1}d_{\gamma}(f_{n}^{(j,k)},
h_{\rho_{jk}}), \\
\nonumber
d_{\gamma}(f_{n}^{(j,k)}, h_{\rho_{jk}}) = -\frac{1}{\gamma}\log\sum_{i=1}^{n}\exp\bigg\{-\frac{\gamma}{2(1-\rho_{jk}^{2})}(Z_{ij}^{2} + Z_{ik}^{2} - 2\rho_{jk}Z_{ij}Z_{ik})\bigg\} + \frac{1}{2(1+\gamma)}\log |1-\rho_{jk}^{2}|.
\end{gather}
The required univariate optimization problem can be solved with
standard techniques.  We provide a projected gradient descent
algorithm in \ref{appendix}.  Finally, we obtain the estimator of
$\bm{\Sigma}_{jk}$ as
$\hat{\bm{\Sigma}}_{jk} =
\sqrt{\hat{\sigma}_{jj}}\sqrt{\hat{\sigma}_{kk}}\hat{\rho}_{jk}$.

\subsection{Existing works}\label{sec3.2}

Some other estimators of $\bm{\Sigma}$ have been proposed under the
cell-wise contamination.  \citet{ollerer15} and \citet{loh15}
considered use of rank correlations.  Based on the decomposition
(\ref{deco}), \citet{loh15} estimated the scale by MAD, and used the
Kendall's tau and Spearman's rho to estimate the correlation.
\citet{ollerer15} proposed to use $Q_{n}$ from \citet{rousseeuw93} for
the scale, and estimate the correlation by the Gaussian rank
correlation from \citet{boudt12}.  \citet{tarr16} directly estimate
$\bm{\Sigma}_{jk}$ following the pairwise approach of
\citet{gnanadesikan72}, which is based on the identity
\begin{align}
  \label{eq:pairwise}
\bm{\Sigma}_{jk} = {\rm Cov}(Y_{j}, Y_{k}) = \frac{1}{4\alpha_{j}\alpha_{k}}
\big\{{\rm Var}(\alpha_{j}Y_{j} + \alpha_{k}Y_{k}) - {\rm Var}(\alpha_{j}Y_{j} - \alpha_{k}Y_{k})\big\},
\end{align}
where $\alpha_{j}=1/\sqrt{{\rm Var}(Y_{j})}$.
\citet{tarr16} proposed to estimate the population variance from the contaminated data using
robust scales such as $Q_{n}$, the $\tau$-scale of \citet{maronna02},  and $P_{n}$ from \citet{tarr12}.

\subsection{Projection of covariance matrix estimate}
We cannot directly plug an estimate of $\bm{\Sigma}$ into the methods
introduced in Section \ref{sec2.1} if it is not positive semidefinite.
The node-wise regression and Glasso require a positive semidefinite
estimate, and CLIME needs a positive definite one.  The estimate
proposed by \citet{ollerer15} is always positive semidefinite, but
this may not be true for the other estimates including the one we
proposed.  However, if an estimate $\hat{\bm{\Sigma}}$ is not positive
semidefinite, we may project it onto the set of positive
(semi)definite matrices.  Different approaches to this projection have been
considered \citep{zhao14}.  We will simply proceed
by solving the problem 
\begin{align*}
\min_{\bm{S}}\|\bm{S} - \hat{\bm{\Sigma}}\|_{F}
\; {\rm\ subject\ to\ } \bm{S} \ge \delta \bm{I}_{p},
\end{align*}
where $\delta \ge 0$, $\|\cdot\|_{F}$ denotes the Frobenius norm and
$\bm{S} \ge \delta \bm{I}_{p}$ means that $\bm{S} - \delta \bm{I}_{p}$
is positive semidefinite.  The solution can be calculated by applying
the singular value decomposition to $\hat{\bm{\Sigma}}$ and then
replacing the singular values $\lambda_{j}$ by
$\max(\lambda_{j}, \delta)$ for each $j=1,\dots, p$.  We set
$\delta=0$ throughout the paper.

\begin{figure}[ht]
\begin{center}
	\includegraphics[width=\linewidth]{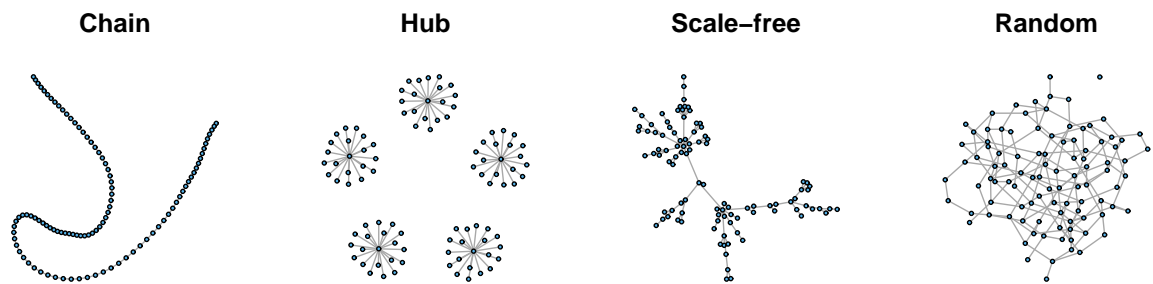}
\end{center}
\vspace{-40pt}
\caption{Four types of true graphs.}
\label{fig1}
\end{figure}

\section{Empirical results}\label{sec4}

\subsection{Simulations}\label{sec4.1}
We provide some simulation results to illustrate the effectiveness of
our method for the graph estimation problem.\footnote{R code to
  implement our method is available at
  \url{https://github.com/shkatayama/robust_graphical_model}.}  We
generated 200 observations from the cell-wise contamination model
(\ref{cmodel}) with $p=100$, $\bm{\mu}=\bm{0}$ and the equal
contamination level
$\varepsilon:=\varepsilon_{1}=\dots = \varepsilon_{p}$.  Both
asymmetric and symmetric contaminations are considered, that is,
$100\varepsilon\%$ of observations in each variable are corrupted by
samples from $N(10, 1)$ in the asymmetric scenario, while half of the
corruption are from $N(-10, 1)$ in the other case.  The true
covariance matrix $\bm{\Sigma}$ determines the true graph structure
via $\bm{\Omega}=\bm{\Sigma}^{-1}$.  We considered four types of true
graphs as shown in Figure \ref{fig1}.  The chain, hub and scale-free
graphs were generated by {\tt huge} package \citep{zhao12}, and the
random graph was made as in \citet{liu12}.

\begin{figure}[h]
\begin{center}
	\includegraphics[width=0.99\linewidth]{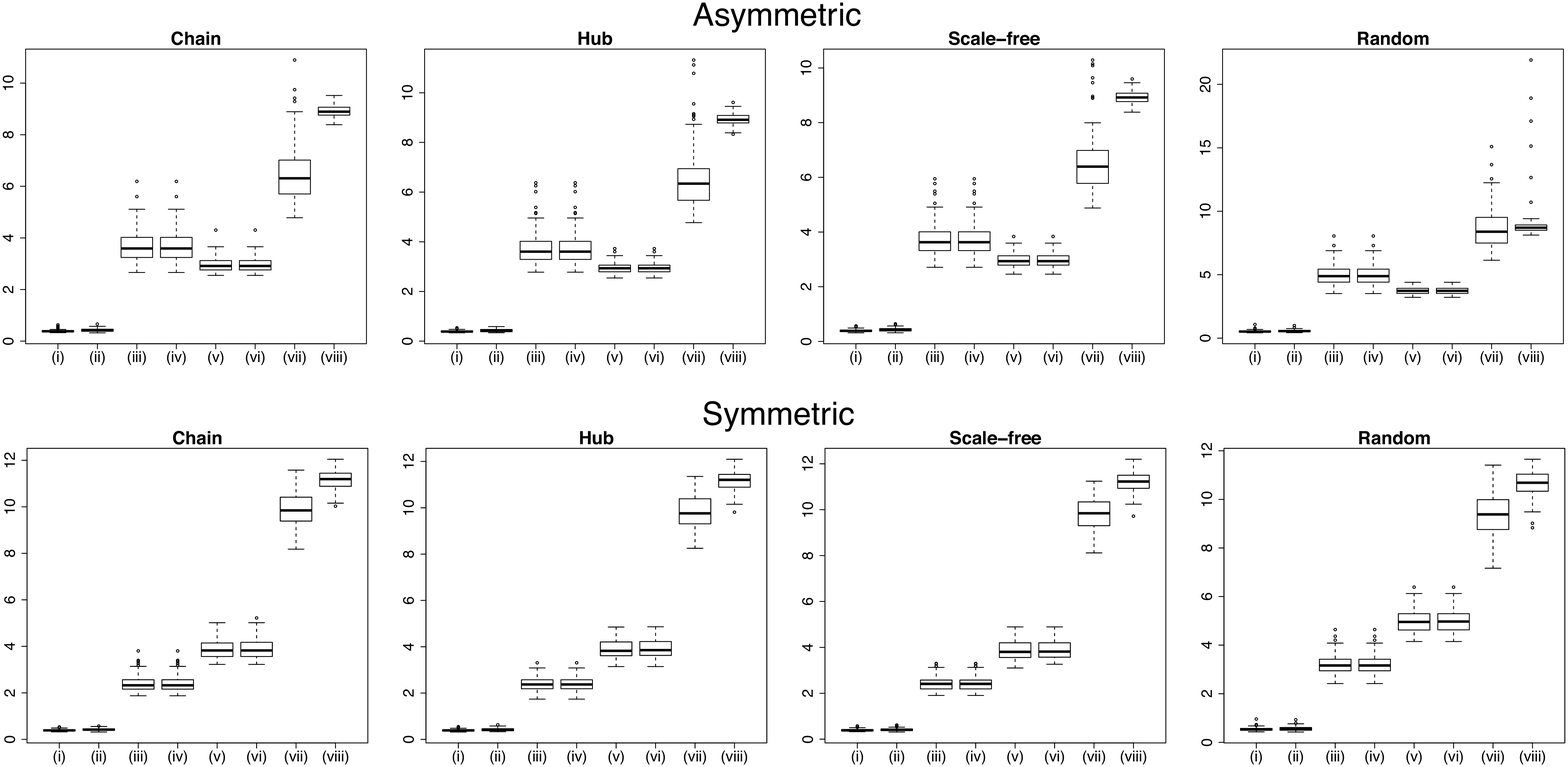}
\end{center}
\caption{Boxplots for our estimators with (i) $\gamma=0.3$ and (ii) $\gamma=0.5$,
(iii) Kendall's tau, (iv) Spearman's rho, (v) Gaussian rank, and pairwise approach with (vi) $Q_{n}$, (vii) $\tau$-scale and
(viii) $P_{n}$ (adaptively trimmed version) when $\varepsilon = 0.25$. 
}
\label{fig2}
\end{figure}

First, we compare our estimator to the existing ones that we described
in Section \ref{sec3.2}.  Performance is evaluated using the measure
$\|\hat{\bm{\Sigma}} - \bm{\Sigma}\|_{\infty}$ that the accuracy of
the resulting graph depends on \citep{jeng11,ravikumar11,cai11}.
Figure \ref{fig2} shows boxplots based on 100 simulations
with $\varepsilon = 0.25$.  Our estimator can be seen to greatly
outperform the others in all cases.  Kendall's tau and Spearman's rho
prefer the symmetric contaminations, while the opposite holds for the
Gaussian rank.  The pairwise approach performs poorly unless $Q_{n}$ is
used.

\begin{figure}[p]
\begin{center}
	\includegraphics[width=0.85\linewidth]{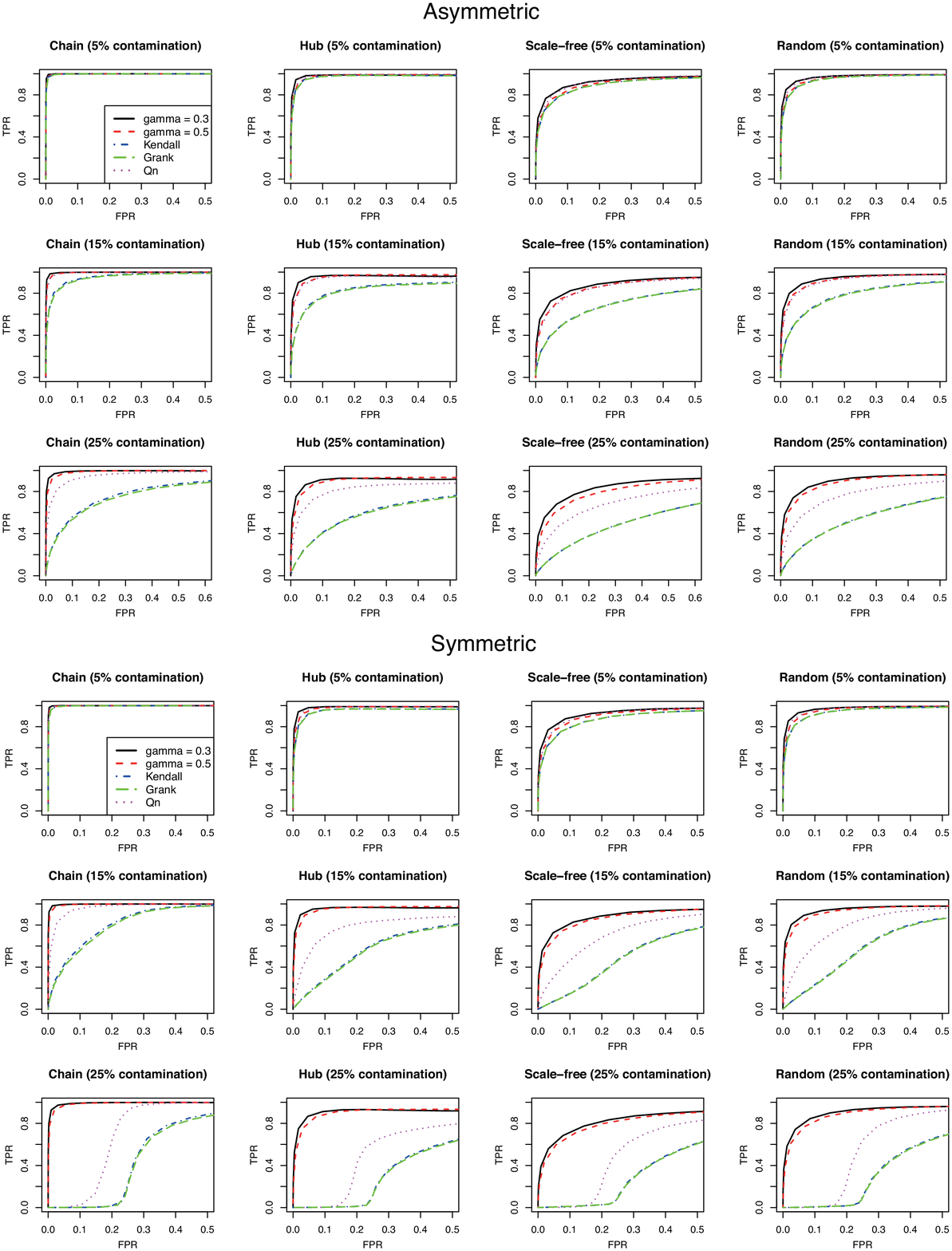}
\end{center}
\caption{ROC curves for Glasso based on our estimators with $\gamma = 0.3$ and $\gamma=0.5$,
Kendall's tau, Gaussian rank and pairwise approach with $Q_{n}$,
for asymmetric and symmetric contaminations at the different levels ($\varepsilon = 0.05, 0.15, 0.25$).}
\label{fig3}
\end{figure}

Next, we consider graph estimation.  We focus on the better performing
competitors, namely, Kendall's tau, Gaussian rank and $Q_{n}$.
Spearman's rho is omitted as it behaved similarly to Kendall's tau.
Furthermore, we restrict attention to Glasso---the other methods are
discussed in the supplement, with similar conclusions.  The Glasso was
implemented using the {\tt QUIC} package of \citet{hsieh11}.
Figure \ref{fig3} illustrates edge recovery via an averaged ROC curve
from 100 simulations.  Each individual ROC curve is a plot of
$({\rm FPR}(\lambda), {\rm TPR}(\lambda))$ versus the tuning parameter
$\lambda$ in~(\ref{eq:glasso}).  Here, for the estimated edge set
$\hat{E} = \hat{E}(\lambda)$ and the true edge set
$E=\{(i,j):\bm{\Omega}_{ij} \neq 0\}$,
${\rm FPR}(\lambda)=|\hat{E}\cap E^{c}|/|E^{c}|$ and
${\rm TPR}(\lambda) = |\hat{E}^{}\cap E|/|E|$.  Inspecting Figure
\ref{fig3}, we can see that our method strongly outperforms the
competitors for all graphs considered, particularly when observations are
highly and symmetrically corrupted.  It is also shown that the
recovery performance of our method hardly changes as the
contamination level increases ($\varepsilon=0.05,0.15,0.25$).

In order to realize the strengths of our method in practice, a
specific value of the tuning parameter $\lambda$ in Glasso needs to be
selected.  We studied this for a 2-fold cross validation approach in
which the observations are randomly split into two folds with nearly equal
size.  A robust covariance matrix $\hat{\bm{\Sigma}}_{k}$ is
calculated on each fold $k=1,2$.  Let
\begin{align*}
L(\lambda)={\rm tr}\left(\hat{\bm{\Sigma}}_{2}\hat{\bm{\Omega}}_{1}(\lambda)\right) - \log \det \hat{\bm{\Omega}}_{1}(\lambda)
\end{align*}
be the negative log-likelihood with $\hat{\bm{\Omega}}_{1}(\lambda)$
estimated only from $\hat{\bm{\Sigma}}_{1}$.  We then select the
tuning parameter by minimizing $L(\lambda)$ over a grid of choices for
$\lambda$.  The main reason for the small number of folds
is that the $\gamma$-divergence needs a sufficient sample size in each
fold for the convergence
$d_{\gamma}(f_{n}, g_{\theta}) \to d_{\gamma}(f,g_{\theta})$ to hold;
recall Section \ref{sec2.2}.  \citet{bickel08} justify the
procedure in high-dimensional covariance estimation.

Table \ref{table1} summarizes the performance of Glasso with tuning
parameter selection when $\varepsilon = 0.25$.  Similar experiments
for $\varepsilon = 0.05$ and $\varepsilon = 0.15$ are described in the supplement.  
The grid for $\lambda$ is chosen as 10 equally spaced
values on the log scale between
$\lambda_{\max}=\|\hat{\bm{\Sigma}} - {\rm
  diag}(\hat{\bm{\Sigma}})\|_{\infty}$ and $0.05\lambda_{\max}$.  
Table \ref{table1} reports the mean squared error (MSE) given by
$\|\hat{\bm{\Omega}} - \bm{\Omega}\|_{F}/p$ in addition to TPR and
FPR.  Our method and $Q_{n}$ show high TPR and low FPR, which suggests
that the tuning parameter is appropriately selected.  Compared with
$Q_{n}$, our method has lower FPR while keeping TPR high.  Moreover,
our method entirely outperforms the competitors in MSE.

\begin{table}[ht]
\centering
\caption{Quantitative performance of Glasso based on the 5 methods when $\varepsilon = 0.25$ and the tuning parameter
is selected by 2-fold cross validation. Each value shows the mean (standard deviation) on 100 simulated data sets.}
\label{table1}
\scalebox{0.7}{
\begin{tabular}{cc|ccc|ccc|ccc|ccc}
& & \multicolumn{3}{c|}{Chain} & \multicolumn{3}{c|}{Hub} & \multicolumn{3}{c|}{Scale-free} & \multicolumn{3}{c}{Random} \\ \hline
& & MSE     & TPR    & FPR    & MSE    & TPR    & FPR   & MSE      & TPR      & FPR      & MSE     & TPR     & FPR    \\ \hline
Asym. & $\gamma=0.3$ & 0.074   & 0.992  & 0.106  & 0.084  & 0.903  & 0.092 & 0.053    & 0.584    & 0.044    & 0.049   & 0.753   & 0.063  \\
& & (0.004)   & (0.011)  & (0.025)  & (0.002)  & (0.045)  & (0.024) & (0.002)  & (0.103)  & (0.020)    & (0.005)   & (0.165)   & (0.031)  \\
& $\gamma=0.5$ & 0.085   & 0.972  & 0.064  & 0.089  & 0.843  & 0.063 & 0.055    & 0.458    & 0.029    & 0.051   & 0.692   & 0.045  \\
& & (0.003) & (0.026) & (0.023)  & (0.002)  & (0.083)  & (0.026) & (0.002)  & (0.114)  & (0.017)  & (0.002)   & (0.079)   & (0.021)  \\
& Kendall  & 0.141   & 0.161  & 0.009  & 0.130  & 0.085  & 0.010 & 0.096   & 0.033    & 0.007    & 0.086   & 0.058   & 0.008  \\
&  & (0.001)  & (0.098) & (0.008) & (0.001) & (0.073) & (0.009) & (0.001)   & (0.032)    & (0.007)    & (0.001)   & (0.045)   & (0.008)  \\
& Grank & 0.142   & 0.145  & 0.008  & 0.130  & 0.069  & 0.007 & 0.096    & 0.035    & 0.006    & 0.086   & 0.056   & 0.007  \\
& & (0.001) & (0.097) & (0.008) & (0.001)  & (0.056)  & (0.006) & (0.001)    & (0.036)    & (0.006)    & (0.001)   & (0.047)   & (0.008)  \\
& $Q_n$  & 0.140   & 0.941  & 0.132  & 0.129  & 0.800  & 0.133 & 0.098    & 0.488    & 0.084    & 0.086   & 0.665   & 0.112  \\
& & (0.001) & (0.028)& (0.035)& (0.001)& (0.067)& (0.042) & (0.001) & (0.090) & (0.028)    & (0.001)   & (0.080)   & (0.037)  \\ \hline
Sym.  & $\gamma=0.3$ & 0.046   & 0.821  & 0.079  & 0.084  & 0.907  & 0.094 & 0.053    & 0.595    & 0.047    & 0.046   & 0.825   & 0.078  \\ 
& & (0.002) & (0.038)  & (0.020)  & (0.002)  & (0.037)  & (0.023) & (0.002)  & (0.068)  & (0.018)    & (0.002)   & (0.045)   & (0.020)  \\
& $\gamma=0.5$ & 0.050   & 0.700  & 0.049  & 0.088  & 0.852  & 0.069 & 0.055    & 0.436    & 0.025    & 0.050   & 0.707   & 0.049  \\
& & (0.002)  & (0.082)  & (0.021)  & (0.002)  & (0.075)  & (0.028) & (0.002)  & (0.102)  & (0.014)   & (0.002)   & (0.083)   & (0.022)  \\
& Kendall  & 0.077   & 0.061  & 0.243  & 0.119  & 0.070  & 0.242 & 0.084    & 0.052    & 0.241    & 0.077   & 0.058   & 0.243  \\
&  & (0.001)  & (0.029) & (0.008)  & (0.001)  & (0.042)  & (0.009) & (0.001) & (0.027)  & (0.008)    & (0.001)   & (0.031)   & (0.009)  \\
& Grank & 0.077   & 0.057  & 0.243  & 0.120  & 0.059  & 0.241 & 0.084    & 0.051  & 0.241    & 0.077   & 0.054   & 0.242  \\
&  & (0.001) & (0.027)  & (0.008)  & (0.001)  & (0.040)  & (0.009) & (0.001) & (0.026)  & (0.009)    & (0.001)   & (0.029)   & (0.010)  \\
& $Q_n$ & 0.082   & 0.744  & 0.292  & 0.127  & 0.653  & 0.267 & 0.094    & 0.505    & 0.248    & 0.082   & 0.742   & 0.288  \\
& & (0.002)   & (0.073)  & (0.028)  & (0.001)  & (0.065)  & (0.024) & (0.001)    & (0.096)    & (0.025)    & (0.002)   & (0.065)   & (0.027) 
\end{tabular}
}
\end{table}

\subsection{Real data analysis}\label{sec4.2}
We consider two applications to gene expression data with smaller
dimension and stock data with large dimension.  Both data sets have
heavy tailed distributions in some variables.  The first example, an
Arabidopsis thaliana data set, is from \citet{wille04} with $n=118$
observations for $p=39$ genes.  The 39 genes are divided into the three
groups: 19 relating to the methylerythritol phosphate (MEP) pathway in
the chloroplast, 15 relating to the mevalonate acid (MVA) pathway in
the cytoplasm and 5 in the mitochondria.  A dense network within each
pathway is expected, but several connections between them have also
been reported and discussed in \citet{wille04}.

\begin{figure}[t]
\begin{center}
	\includegraphics[width=0.95\linewidth]{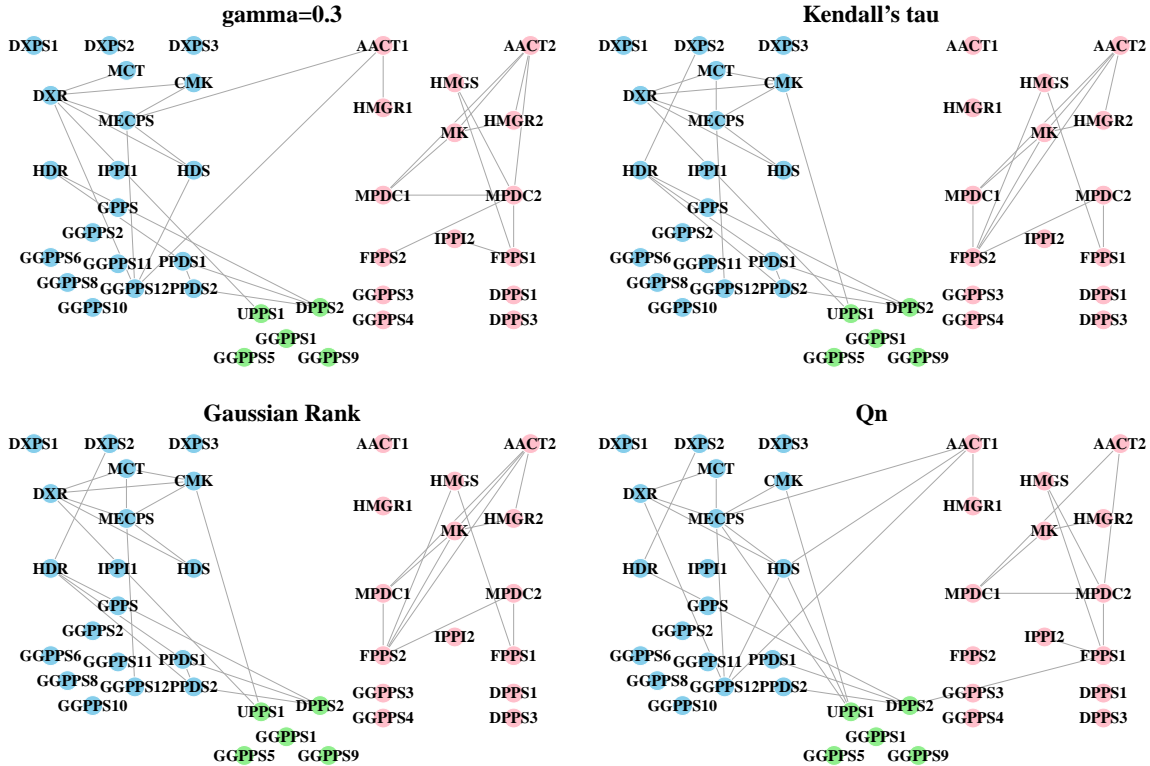}
\end{center}
\caption{Graphs estimated with Glasso based on our estimator
  ($\gamma=0.3$), Kendall's tau, Gaussian Rank and pairwise approach
  with $Q_{n}$ for the Arabidopsis thaliana data set.  Each node
  corresponds to a gene, and each graph has 30 edges.  Genes colored
  blue and red are in the MEP and MVA pathways, respectively.  Mitochondria 
  genes are colored green.}
\label{fig4}
\end{figure}

The estimated graphs are shown in Figure \ref{fig4}.  Before
robust covariance estimation, we standardized the data using median
and MAD.  The tuning parameter of Glasso was selected to obtain 30
edges which is roughly number of edges considered in \citet{wille04}. 
We can see from Figure \ref{fig4} that our method and $Q_{n}$
identify a connection between the MEP and MVA pathways but Kendall's
tau and Gaussian rank do not.  There are slight differences between
our method and $Q_{n}$.  Our method outputs more dense networks within
both MEP and MVA pathways, while $Q_{n}$ connects AACT1 and HDS.  The
two methods agree that AACT1 is the hub connecting the two pathways.
Though \citet{wille04} have reported that HMGR1 is also a hub, if we
trust our robust analysis, HMGR1 may link to the MEP pathway just
through AACT1.

\begin{figure}[ht]
\begin{center}
	\includegraphics[width=0.95\linewidth]{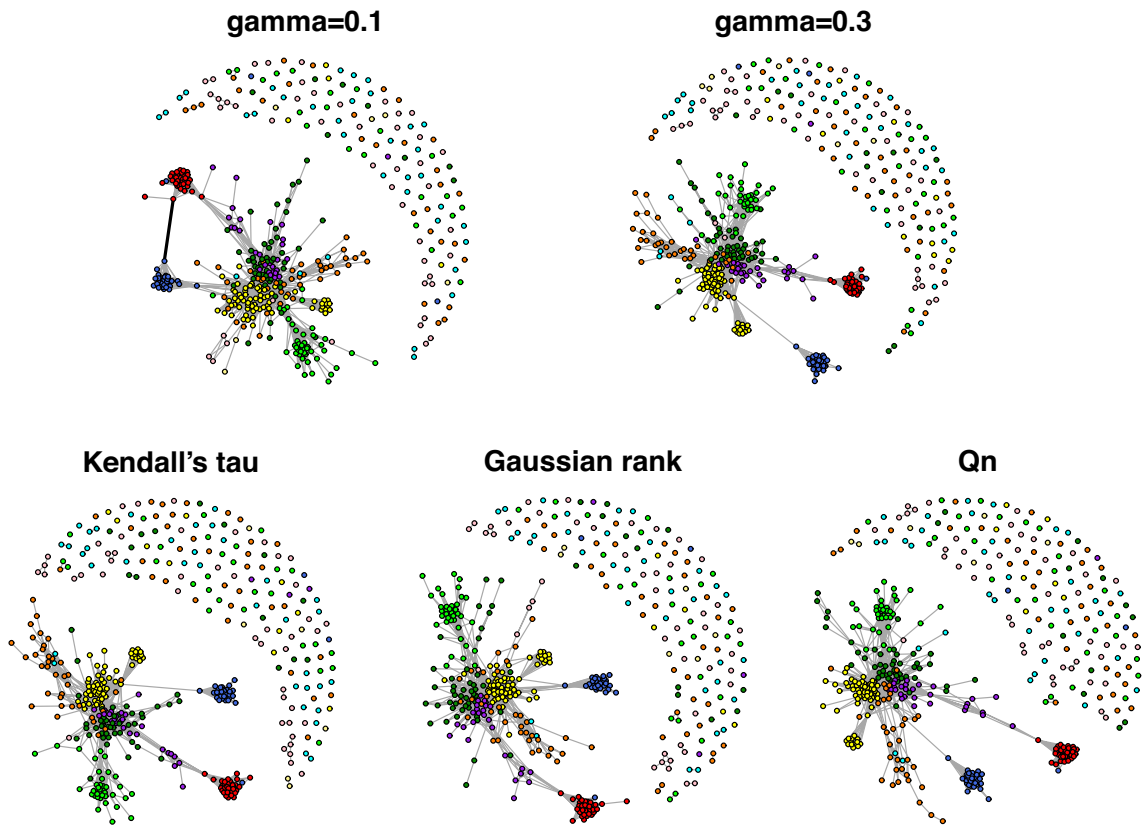}
\end{center}
\caption{Graphs estimated with Glasso based on our estimator
  ($\gamma=0.1,0.3$), Kendall's tau, Gaussian Rank and the pairwise
  approach with $Q_{n}$ for the S\&P 500 stock data set.  Each graph
  has 2,500 edges.  Each node represents a stock, and stocks from the
  same GICS sector have the same color.  A stock in the ``Utilities''
  (blue) sector links to a stock in the ``Materials'' (red) sector
  only for our method with $\gamma = 0.1$.  This edge is drawn bold.  }
\label{fig5}
\end{figure}

The second example is data on the daily closing prices of the S\&P 500
stocks from January 1, 2003 to January 1, 2008.  Preprocessing as in
\citet{zhao12}, there are $n=1257$ observations for $p=452$ stocks.  The
stocks are divided into 10 Global Industry Classification Standard
(GICS) sectors.  We proceeded as in the previous application but
selected the tuning parameter to have a total of 2,500 edges, which results in well-clustered structure.  
Figure \ref{fig5} illustrates the results, now also considering
$\gamma = 0.1$.  Stocks in the same GICS sector are shown in the same
color.  Although the estimated graphs are quite similar, only our
method with $\gamma = 0.1$ identifies a direct connection between a
stock in the ``Utilities'' (blue) sector and a stock in the
``Materials'' (red) sector.  

\section{Concluding remarks}\label{sec5}

We have introduced novel methodology for robust estimation of a
conditional independence graph via $\gamma$-divergence.  The method is
designed for cell-wise contamination and is able to extract available
information from multivariate data even when they are high-dimensional
with corrupted values in many/most observations.  Our method strongly
outperformed competitors in our simulations.  In particular, it showed
very good behavior across different levels of contaminations.

A noteworthy result was found for the pairwise approach with $Q_{n}$;
recall~(\ref{eq:pairwise}).  For asymmetric contamination it performed
well even at high contamination levels, but it performed poorly for
symmetric scenario.  This imbalance can be explained as follows.
For univariate samples $X_{1},\dots,X_{n}$, the $Q_{n}$ is
based on the first quantile of $\{|X_{i} - X_{j}|:i<j\}$.  If both
$X_{i}$ and $X_{j}$ are contamined as a $N(10, 1)$ draw, then the
difference $X_{i} - X_{j} \sim N(0,2)$ behaves as it does for clean
observations.  However, this is not the case for symmetric
contamination with, say, $X_i\sim N(10,1)$ and $X_j\sim N(-10,1)$.

Our experiments in Section~\ref{sec4} show that our method can achieve
good results with a fixed default value for divergence parameter
$\gamma$.  Of course, further improvements are possible by tuning this
parameter.  This, however, is challenging because an optimal choice of
$\gamma$ would depend on the typically unknown contamination density
and level.  If entirely clean sub-samples were available, then
$\gamma$ could be tuned by comparing the sample covariance matrix of
the sub-samples and the robust covariance matrix obtained via
$\gamma$-divergence of the other samples that may include
contaminations.

For the simpler approach that uses rank correlations, \citet{loh15}
were able to give an analysis of the estimation error
$\|\hat{\bm{\Sigma}} - \bm{\Sigma}\|_{\infty}$.  Obtaining analogous
results for the estimator via $\gamma$-divergence is an interesting
open problem for future work.  A key challenge is the non-convexity of
the objective function, which makes the results of \citet{miao10} and
\citet{catoni12} inapplicable.  However, we believe that some
convexity properties hold on a restricted parameter space and may
offer a way to analyze the estimator.

\appendix
\def\thesection{Appendix\;\Alph{section}}
\section{Projected gradient descent algorithm}\label{appendix}
We outline the projected gradient descent algorithm for computation of
$\hat{\rho}_{jk}$ from~(\ref{eq:gamma-rho}).  To avoid numerical
singularity, we replace the restriction $|\rho_{jk}| < 1$ by
$|\rho_{jk}| \le R$ with $R\approx 1$.  For simpler notation, let
$d_{\gamma}(\rho_{jk}) = d_{\gamma}(f_{n}^{(j,k)}, h_{\rho_{jk}})$.
The gradient of this function is
\begin{align*}
\nabla d_{\gamma}(\rho_{jk}) = \frac{\gamma}{(1-\rho_{jk}^{2})^{2}}\sum_{i=1}^{n}w_{ijk}
\big\{(1+\rho_{jk}^{2})Z_{ij}Z_{ik} - \rho_{jk}(Z_{ij}^{2} + Z_{ik}^{2})\big\}-\frac{1}{1+\gamma}\frac{\rho_{jk}}{1-\rho_{jk}^{2}},
\end{align*}
where
\footnotesize
\begin{align*}
w_{ijk}=\exp\bigg\{-\frac{\gamma}{2(1-\rho_{jk}^{2})}\big(Z_{ij}^{2} + Z_{ik}^{2} -2\rho_{jk}Z_{ij}Z_{ik}\big)\bigg\}\bigg/
\sum_{i=1}^{n}\exp\bigg\{-\frac{\gamma}{2(1-\rho_{jk}^{2})}\big(Z_{ij}^{2} + Z_{ik}^{2} -2\rho_{jk}Z_{ij}Z_{ik}\big)\bigg\}.
\end{align*}
\normalsize
The objective function $d_{\gamma}(\rho_{jk})$ is locally approximated around $\rho_{jk}'$ by
\begin{align*}
\phi_{\gamma}(\rho_{jk}; \rho_{jk}') = d_{\gamma}(\rho_{jk}') + \nabla d_{\gamma}(\rho_{jk}')(\rho_{jk} - \rho_{jk}')
+\frac{s}{2}(\rho_{jk} - \rho_{jk}')^{2},
\end{align*}
where $s > 0$ is the step size parameter. We select sufficiently large $s$ such that 
$d_{\gamma}(\rho_{jk}) \le \phi_{\gamma}(\rho_{jk}; \rho_{jk}')$.
The projected gradient descent minimizes $\phi_{\gamma}(\rho_{jk}; \rho_{jk}')$ over $|\rho_{jk}| \le R$
instead of $d_{\gamma}(\rho_{jk})$. 
The minimizer is
${\rm sgn}(\bar{\rho}_{jk}')\min(|\bar{\rho}_{jk}'|, R)$ with
$\bar{\rho}_{jk}' = \rho_{jk}' - s^{-1}\nabla d_{\gamma}(\rho_{jk}')$.
Algorithm \ref{algorithm1} summarizes the procedure.

\begin{algorithm}
\SetAlgoNoLine
\KwIn{
Standardized data $(Z_{1j}, Z_{1k}),\dots, (Z_{nj}, Z_{nk})$; 
divergence parameter $\gamma > 0$.
}
Initialize $t=0$, $\rho_{jk}^{0} = 0$, $s^{0} = 0.01$ and set $R = 0.99$.

\Repeat{convergence}{
$t_{in} \leftarrow 0$

$s^{t,t_{in}} \leftarrow s^{t}$

\Repeat{$d_{\gamma}(\rho_{jk}^{t}) \le \phi_{\gamma}(\rho_{jk}^{t}; \rho_{jk}^{t,t_{in}})$}{
$\nu_{jk}^{t, t_{in}} \leftarrow  \rho_{jk}^{t} - \nabla d_{\gamma}(\rho_{jk}^{t})/s^{t,t_{in}}$

$\rho_{jk}^{t, t_{in}+1} \leftarrow {\rm sgn}(\nu_{jk}^{t,t_{in}})\min(|\nu_{jk}^{t,t_{in}}|, R)$

$s^{t,t_{in}+1} \leftarrow 2s^{t,t_{in}}$ 

$t_{in} \leftarrow t_{in} + 1$
}

$\rho_{jk}^{t+1} \leftarrow \rho_{jk}^{t,t_{in}}$

$s^{t+1} \leftarrow s^{t,t_{in}}/2$

$t\leftarrow t + 1$
}
\caption{Projected gradient descent algorithm for $\hat{\rho}_{jk}$}
\label{algorithm1}
\end{algorithm}

\end{document}


\maketitle
\footnotesize 
\begin{center}
$^1$Department of Industrial Engineering and Economics, Tokyo Institute of Technology, Japan \\ 
$^2$The Institute of Statistical Mathematics, Japan \\
$^3$Nagoya University Graduate School of Medicine, Japan \\
$^4$Department of Statistics, University of Washington, USA

\end{center}
\normalsize

\appendix
\renewcommand{\thesection}{S\arabic{section}}
\section{Additional simulations}
\renewcommand{\thefigure}{S\arabic{figure}}
\renewcommand{\thetable}{S\arabic{table}}

This supplementary article provides additional simulation results.
The data generating process is the same as the article, except that we
restrict the dimension to $p=50$ for the CLIME simulation due to long
computation time.  The node-wise regression was carried out using the
{\tt huge} package of \citet{zhao12} and CLIME was computed with the
{\tt clime} package of \citet{cai11}.  We used the perturbation
default set in the {\tt clime} package.  Negative definite robust
covariance matrices were projected to positive semidefiniteness using the procedure from the main article.

\subsection{ROC curves}

\begin{figure}[p]
\begin{center}
	\includegraphics[width=0.85\linewidth]{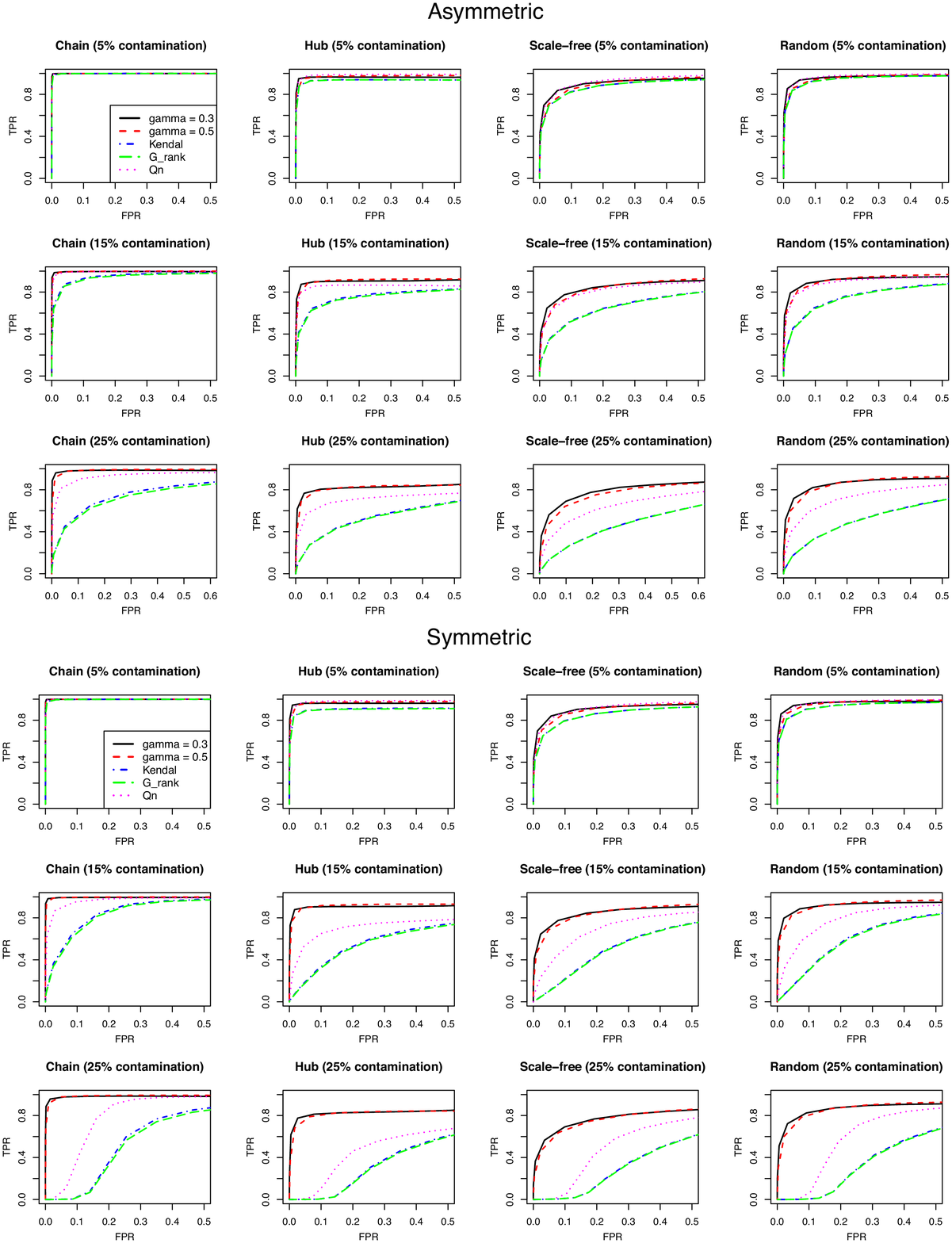}
\end{center}
\caption{ROC curves for node-wise regression based on our estimators with $\gamma = 0.3$ and $\gamma=0.5$,
Kendall's tau, Gaussian rank and pairwise approach with $Q_{n}$,
for asymmetric and symmetric contaminations at the different levels ($\varepsilon = 0.05, 0.15, 0.25$).}
\label{figS1}
\end{figure}

\begin{figure}[p]
\begin{center}
	\includegraphics[width=0.85\linewidth]{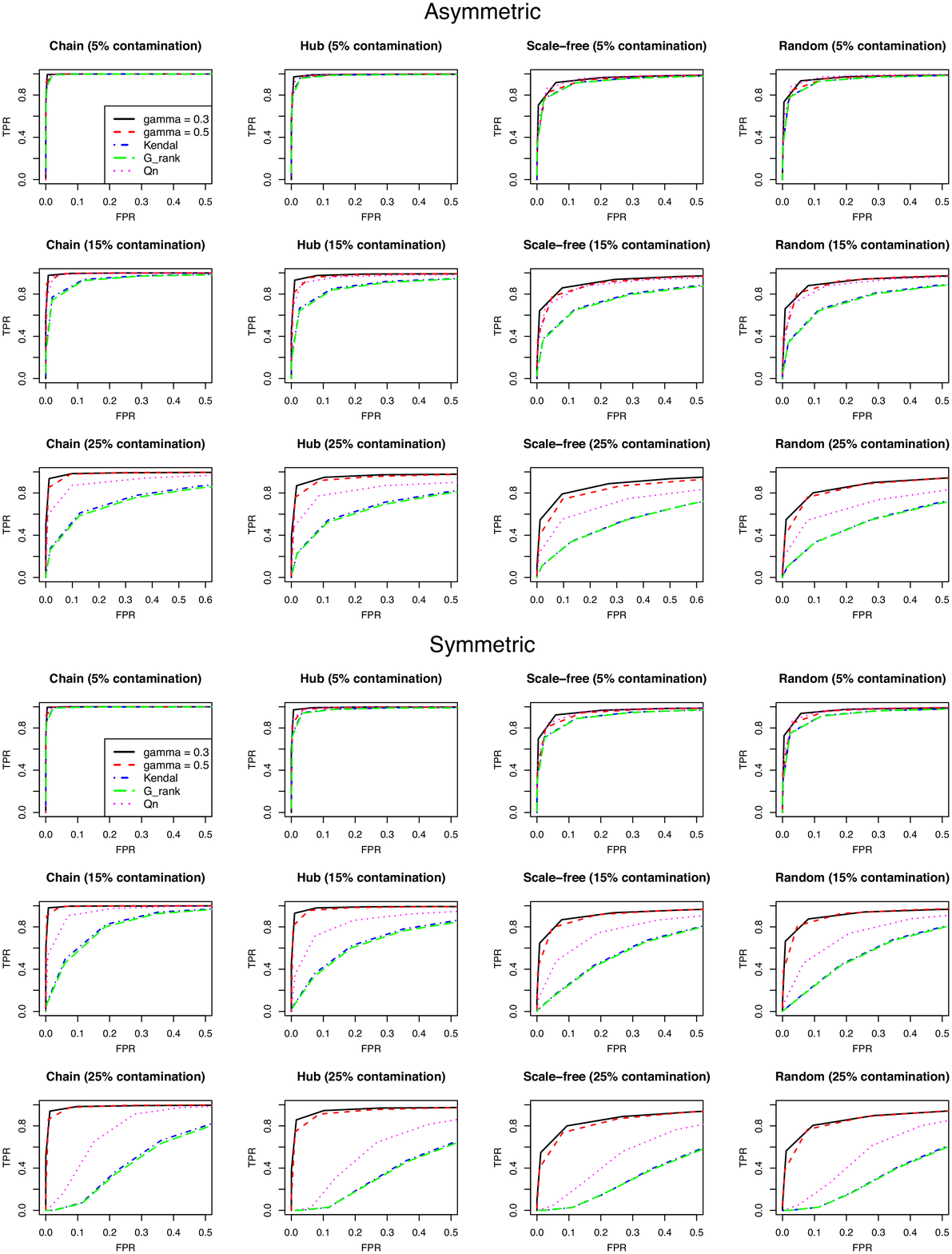}
\end{center}
\caption{ROC curves for CLIME based on our estimators with $\gamma = 0.3$ and $\gamma=0.5$,
Kendall's tau, Gaussian rank and pairwise approach with $Q_{n}$,
for asymmetric and symmetric contaminations at the different levels ($\varepsilon = 0.05, 0.15, 0.25$).}
\label{figS2}
\end{figure}

The ROC curves for the node-wise regression and CLIME are provided in
Figure \ref{figS1} and Figure \ref{figS2}.  We observe behavior
similarly to that discussed for the Glasso in the main article.

\subsection{Quantitive performances}

Tables \ref{tableS1} and \ref{tableS2} summarize the performance of
Glasso with the tuning parameter selected by the 2-fold cross
validation when $\varepsilon=0.05$ and $\varepsilon = 0.15$,
respectively.  The result for $\varepsilon = 0.25$ is reported in the
article.  For the low contamination level $\varepsilon = 0.05$, the
five methods are comparable, but our method outperforms the competitors
for higher contamination level.  Tables \ref{tableS3}--\ref{tableS5}
show the results of node-wise regression.  The edge set was estimated
by the "OR" rule, and the tuning parameter was selected by the
Stability Approach for Regularization Selection (StARS,
\citeauthor{liu10}, \citeyear{liu10})
with 10 times sub-sampling to size $n/2$ and the cut point value
$0.2$.  Tables \ref{tableS6}--\ref{tableS8} provide the results
of CLIME with the tuning parameter selected by 2-fold cross
validation.  With both node-wise regression and CLIME, we see that
the tuning parameter selection does not work well for $Q_{n}$ when
$\varepsilon=0.25$, in contrast to the case of Glasso.

\begin{table}[p]
\centering
\caption{Quantitative performances of Glasso based on the 5 methods when $\varepsilon = 0.05$ and the tuning parameter
is selected by 2-fold cross validation. Each value shows the mean (standard deviation) on 100 simulated data set.}
\label{tableS1}
\scalebox{0.7}{
\begin{tabular}{cc|ccc|ccc|ccc|ccc}
& & \multicolumn{3}{c|}{Chain} & \multicolumn{3}{c|}{Hub} & \multicolumn{3}{c|}{Scale-free} & \multicolumn{3}{c}{Random} \\ \hline
& & MSE & TPR & FPR & MSE & TPR & FPR & MSE  & TPR  & FPR  & MSE & TPR & FPR    \\ \hline
Asym. & $\gamma=0.3$ & 0.066 & 1.000 & 0.088 & 0.075& 0.987& 0.084 & 0.048 & 0.788 & 0.041& 0.041& 0.947 & 0.071  \\
& & (0.003) & (0.002)& (0.017)& (0.002)  & (0.012)  & (0.020) & (0.002)  & (0.067)  & (0.018)  & (0.002) & (0.023)   & (0.025)  \\
& $\gamma=0.5$ & 0.080   & 0.998  & 0.058  & 0.083  & 0.973  & 0.075 & 0.052  & 0.664  & 0.028  & 0.046 & 0.892   & 0.057  \\
&  & (0.003) & (0.004) & (0.026)  & (0.002)  & (0.023)  & (0.025) & (0.002)  & (0.093)  & (0.014)  & (0.002)   & (0.054) & (0.021  \\
& Kend  & 0.086   & 0.998  & 0.076  & 0.088  & 0.969  & 0.084 & 0.058  & 0.654  & 0.034    & 0.052   & 0.892   & 0.066  \\
&  & (0.004) & (0.004)  & (0.031) & (0.003)  & (0.028)  & (0.025) & (0.002)  & (0.086)  & (0.016)  & (0.002) & (0.047)   & (0.024)  \\
& Grank & 0.088   & 0.997  & 0.067  & 0.089  & 0.963  & 0.078 & 0.059  & 0.638  & 0.032    & 0.052   & 0.877   & 0.059  \\
& & (0.004) & (0.007)  & (0.030)  & (0.003)  & (0.030)  & (0.023)& (0.002)  & (0.095)  & (0.016)  & (0.002)   & (0.052) & (0.024)  \\
& $Q_{n}$ & 0.097 & 0.999  & 0.059  & 0.095  & 0.982  & 0.076 & 0.062  & 0.689  & 0.027    & 0.056   & 0.925   & 0.057  \\
&  & (0.002) & (0.003)  & (0.026)  & (0.002)  & (0.021)  & (0.026) & (0.001)  & (0.105) & (0.014) & (0.001) & (0.034) & (0.019)  \\ \hline
Sym.  & $\gamma=0.3$ & 0.066 & 1.000  & 0.087  & 0.076  & 0.984 & 0.083 & 0.048 & 0.786 & 0.039 & 0.041 & 0.948 & 0.070  \\
&  & (0.003) & (0.002) & (0.016)  & (0.002)  & (0.014)  & (0.022) & (0.002) & (0.062) & (0.017) & (0.002) & (0.021) & (0.023)  \\
& $\gamma=0.5$ & 0.080 & 0.999  & 0.059  & 0.083  & 0.969  & 0.074 & 0.052  & 0.661  & 0.026  & 0.046 & 0.888 & 0.055  \\
&  & (0.003) & (0.004)  & (0.025)  & (0.002)  & (0.029)  & (0.025) & (0.002)  & (0.095)  & (0.013)    & (0.002)   & (0.063)   & (0.025)  \\
& Kend  & 0.088   & 0.998  & 0.071  & 0.090  & 0.949  & 0.088 & 0.059    & 0.647    & 0.036    & 0.052   & 0.867   & 0.066  \\
&  & (0.004)  & (0.005)  & (0.030)  & (0.003)  & (0.033)  & (0.026) & (0.002)    & (0.097) & (0.016)  & (0.002)   & (0.056)  & (0.026)  \\
& Grank & 0.089   & 0.996  & 0.064  & 0.090  & 0.940  & 0.080 & 0.059    & 0.624    & 0.033    & 0.052   & 0.862   & 0.063  \\
& & (0.004)  & (0.006)  & (0.027)  & (0.003)  & (0.042)  & (0.028) & (0.002)  & (0.106)  & (0.016)  & (0.002) & (0.056)   & (0.024)  \\
& $Q_n$ & 0.097   & 0.999  & 0.060  & 0.095  & 0.979  & 0.086 & 0.062    & 0.699    & 0.031  & 0.056   & 0.905   & 0.055  \\
& & (0.002)   & (0.003)  & (0.026)  & (0.002)  & (0.022)  & (0.025) & (0.001)  & (0.091) & (0.014)& (0.001)   & (0.051)   & (0.022) 
\end{tabular}
}
\end{table}

\begin{table}[p]
\centering
\caption{Quantitative performance of Glasso based on the 5 methods when $\varepsilon = 0.15$ and the tuning parameter
is selected by 2-fold cross validation. Each value shows the mean (standard deviation) on 100 simulated data sets.}
\label{tableS2}
\scalebox{0.7}{
\begin{tabular}{cc|ccc|ccc|ccc|ccc}
&  & \multicolumn{3}{c|}{Chain} & \multicolumn{3}{c|}{Hub} & \multicolumn{3}{c|}{Scale-free} & \multicolumn{3}{c}{Random} \\ \hline
&  & MSE  & TPR    & FPR    & MSE    & TPR    & FPR   & MSE      & TPR      & FPR      & MSE     & TPR     & FPR    \\ \hline
Asym. & $\gamma=0.3$ & 0.068 & 0.999  & 0.099  & 0.079  & 0.957  & 0.087 & 0.051 & 0.681& 0.040  & 0.043   & 0.894   & 0.073  \\
& & (0.003)   & (0.004)  & (0.019)  & (0.002)  & (0.022)  & (0.022) & (0.002)  & (0.075)  & (0.018)    & (0.002)   & (0.031) & (0.021)  \\
&  $\gamma=0.5$ & 0.083   & 0.993  & 0.056  & 0.087  & 0.917  & 0.060 & 0.054    & 0.520    & 0.022    & 0.049   & 0.783   & 0.041  \\
&  & (0.003)   & (0.010)  & (0.022)  & (0.002)  & (0.048)  & (0.024) & (0.002)    & (0.113)  & (0.014) & (0.002)   & (0.072) & (0.020) \\
& Kend  & 0.120   & 0.832  & 0.037  & 0.113  & 0.564  & 0.034 & 0.078    & 0.226    & 0.016    & 0.071   & 0.387   & 0.023  \\
&  & (0.002)   & (0.076)  & (0.01)5  & (0.002)  & (0.144)  & (0.021) & (0.002) & (0.092)  & (0.011)  & (0.001)   & (0.107)   & (0.015)  \\
& Grank & 0.120   & 0.805  & 0.035  & 0.113  & 0.555  & 0.033 & 0.078    & 0.203    & 0.013    & 0.072   & 0.361   & 0.020  \\
&  & (0.002)   & (0.084)  & (0.015)  & (0.002)  & (0.136)  & (0.019) & (0.002) & (0.094) & (0.010)    & (0.001)   & (0.110)   & (0.014)  \\
& $Q_n$    & 0.119   & 0.996  & 0.094  & 0.114  & 0.929  & 0.082 & 0.083    & 0.613    & 0.041 & 0.073   & 0.842   & 0.074  \\
& & (0.002) & (0.006)  & (0.030)  & (0.002)  & (0.034)  & (0.026) & (0.001) & (0.087)  & (0.018)  & (0.001) & (0.057) & (0.025)  \\ \hline
Sym.  & $\gamma=0.3$ & 0.069   & 0.998  & 0.101  & 0.079  & 0.961  & 0.093 & 0.050    & 0.704    & 0.044    & 0.043   & 0.899   & 0.076  \\
& & (0.002)  & (0.004) & (0.018)  & (0.002)  & (0.020)  & (0.021) & (0.002)    & (0.073)    & (0.020)   & (0.002)   & (0.032)   & (0.021)  \\
& $\gamma=0.5$ & 0.083   & 0.992  & 0.053  & 0.087  & 0.910  & 0.059 & 0.054    & 0.546    & 0.024    & 0.049 & 0.783   & 0.043  \\
& & (0.002)   & (0.010)  & (0.018)  & (0.002)  & (0.053)  & (0.025) & (0.002)    & (0.114)    & (0.015)    & (0.002)   & (0.076)& (0.022)  \\
& Kend  & 0.117   & 0.698  & 0.141 & 0.110& 0.416  & 0.155 & 0.074    & 0.177    & 0.113    & 0.068   & 0.283   & 0.124  \\
& & (0.002   & (0.083)  & (0.034)& (0.001)  & (0.080)  & (0.026) & (0.001) & (0.069)    & (0.035) & (0.001) & (0.076)   & (0.031)  \\
& Grank & 0.117   & 0.648  & 0.130  & 0.110  & 0.385  & 0.148 & 0.074    & 0.166    & 0.111 & 0.068   & 0.269   & 0.123  \\
& & (0.002)   & (0.088)  & (0.032)  & (0.001)  & (0.085)  & (0.026) & (0.001)    & (0.065)    & (0.036) & (0.001)   & (0.068)   & (0.030)  \\
& $Q_n$  & 0.120   & 0.988  & 0.199  & 0.116  & 0.791  & 0.173 & 0.083    & 0.501    & 0.116    & 0.074   & 0.748   & 0.170  \\
&  & (0.002)   & (0.012) & (0.033)  & (0.001)  & (0.051)  & (0.025) & (0.001)    & (0.097)    & (0.035)    & (0.001)   & (0.057)  & (0.031) 
\end{tabular}
}
\end{table}

\begin{table}[p]
\centering
\caption{Quantitative performance of node-wise regression based on the 5 methods 
when $\varepsilon = 0.05$ and the tuning parameter
is selected by StARS. Each value shows the mean (standard deviation) on 100 simulated data sets.}
\label{tableS3}
\scalebox{0.8}{
\begin{tabular}{cc|cc|cc|cc|cc}
&& \multicolumn{2}{c|}{Chain} & \multicolumn{2}{c|}{Hub} & \multicolumn{2}{c|}{Scale-free} & \multicolumn{2}{c}{Random} \\ \hline
&& TPR  & FPR         & TPR        & FPR        & TPR            & FPR           & TPR          & FPR         \\ \hline
Asym. & $\gamma=0.3$ & 1.000       & 0.059       & 0.958      & 0.053      & 0.805          & 0.042         & 0.936        & 0.053       \\
           &       & (0.002)       & (0.018)       & (0.022)      & (0.015)      & (0.056)          & (0.019)         & (0.027)        & (0.016)       \\
           & $\gamma=0.5$ & 0.998       & 0.028       & 0.949      & 0.038      & 0.712          & 0.032         & 0.848        & 0.028       \\
           &       & (0.005)      & (0.012)       & (0.031)      & (0.016)      & (0.076)          & (0.014)         & (0.047)        & (0.014)      \\
           & Kend  & 0.996       & 0.038       & 0.921      & 0.046      & 0.684          & 0.032         & 0.868        & 0.042       \\
           &       & (0.006)       & (0.019)       & (0.029)      & (0.017)      & (0.080)          & (0.017)         & (0.046)        & (0.016)       \\
           & Grank & 0.996       & 0.037       & 0.920      & 0.045      & 0.658          & 0.028         & 0.862        & 0.041       \\
           &       & (0.006)       & (0.020)       & (0.031)      & (0.017)      & (0.106)          & (0.016)         & (0.048)        & (0.016)       \\
           & $Q_n$    & 0.999       & 0.036       & 0.975      & 0.036      & 0.747          & 0.029         & 0.894        & 0.028       \\
           &       & (0.002)       & (0.012)       & (0.017)      & (0.015)      & (0.082)          & (0.013)         & (0.039)        & (0.011)       \\ \hline
Sym.  & $\gamma=0.3$ & 1.000       & 0.056       & 0.962      & 0.051      & 0.797          & 0.041         & 0.934        & 0.049       \\
           &       & (0.002)       & (0.018)       & (0.018)      & (0.015)      & (0.063)          & (0.019)         & (0.025)        & (0.016)       \\
           & $\gamma=0.5$ & 0.998       & 0.025       & 0.958      & 0.041      & 0.685          & 0.026         & 0.862        & 0.030       \\
           &       & (0.006)       & (0.011)       & (0.025)      & (0.017)      & (0.074)          & (0.012)         & (0.050)        & (0.014)       \\
           & Kend  & 0.995       & 0.042       & 0.895      & 0.050      & 0.649          & 0.034         & 0.827        & 0.041       \\
           &       & (0.008)       & (0.021)       & (0.039)      & (0.016)      & (0.072)          & (0.015)         & (0.053)        & (0.019)       \\
           & Grank & 0.994       & 0.039       & 0.889      & 0.047      & 0.636          & 0.031         & 0.814        & 0.038       \\
           &       & (0.009)       & (0.021)       & (0.042)      & (0.015)      & (0.078)          & (0.014)         & (0.059)        & (0.018)      \\
           & $Q_n$    & 0.999       & 0.035       & 0.973      & 0.039      & 0.714          & 0.028         & 0.882        & 0.030       \\
           &       & (0.003)       & (0.011)       & (0.017)      & (0.015)      & (0.084)          & (0.014)         & (0.052)        & (0.013)      
\end{tabular}
}
\end{table}

\begin{table}[p]
\centering
\caption{Quantitative performance of node-wise regression based on the 5 methods 
when $\varepsilon = 0.15$ and the tuning parameter
is selected by StARS. Each value shows the mean (standard deviation) on 100 simulated data sets.}
\label{tableS4}
\scalebox{0.8}{
\begin{tabular}{cc|cc|cc|cc|cc}
& & \multicolumn{2}{c|}{Chain} & \multicolumn{2}{c|}{Hub} & \multicolumn{2}{c|}{Scale-free} & \multicolumn{2}{c}{Random} \\ \hline
& & TPR         & FPR         & TPR        & FPR        & TPR            & FPR           & TPR          & FPR         \\ \hline
Asym. & $\gamma=0.3$ & 0.993       & 0.065       & 0.887      & 0.064      & 0.722          & 0.048         & 0.859        & 0.055       \\
           &       & (0.009)       & (0.024)       & (0.029)      & (0.017)      & (0.067)          & (0.019)         & (0.044)        & (0.021)       \\
           & $\gamma=0.5$ & 0.992       & 0.035       & 0.872      & 0.039      & 0.604          & 0.031         & 0.749        & 0.028       \\
           &       & (0.008)       & (0.013)       & (0.038)      & (0.016)      & (0.080)          & (0.014)         & (0.058)        & (0.012)       \\
           & Kend  & 0.849       & 0.035       & 0.610      & 0.039      & 0.355          & 0.033         & 0.462        & 0.034       \\
           &       & (0.057)       & (0.018)       & (0.085)      & (0.014)      & (0.092)          & (0.016)         & (0.079)        & (0.015)       \\
           & Grank & 0.818       & 0.031       & 0.597      & 0.039      & 0.341          & 0.031         & 0.441        & 0.031       \\
           &       & (0.066)       & (0.017)       & (0.085)      & (0.013)      & (0.094)          & (0.015)         & (0.085)        & (0.014)       \\
           & $Q_n$    & 0.989       & 0.046       & 0.855      & 0.049      & 0.636          & 0.042         & 0.765        & 0.038       \\
           &       & (0.012)       & (0.018)       & (0.040)      & (0.019)      & (0.067)          & (0.017)         & (0.058)        & (0.018)       \\ \hline
Sym.  & $\gamma=0.3$ & 0.995       & 0.058       & 0.895      & 0.064      & 0.725          & 0.052         & 0.869        & 0.056       \\
           &       & (0.007)       & (0.024)       & (0.026)      & (0.017)      & (0.073)          & (0.021)         & (0.034)        & (0.019)       \\
           & $\gamma=0.5$ & 0.990       & 0.034       & 0.881      & 0.038      & 0.584          & 0.027         & 0.778        & 0.032       \\
           &       & (0.009)       & (0.014)       & (0.035)      & (0.016)      & (0.078)          & (0.012)         & (0.058)        & (0.014)       \\
           & Kend  & 0.626       & 0.079       & 0.285      & 0.082      & 0.151          & 0.068         & 0.203        & 0.067       \\
           &       & (0.077)       & (0.018)       & (0.071)      & (0.017)      & (0.064)          & (0.023)         & (0.075)        & (0.023)       \\
           & Grank & 0.583       & 0.075       & 0.278      & 0.083      & 0.143          & 0.065         & 0.189        & 0.064       \\
           &       & (0.084)       & (0.019)       & (0.072)      & (0.017)      & (0.063)          & (0.023)         & (0.077)        & (0.023)       \\
           & $Q_n$    & 0.950       & 0.091       & 0.623      & 0.083      & 0.425          & 0.067         & 0.561        & 0.073       \\
           &       & (0.029)       & (0.023)       & (0.060)      & (0.021)      & (0.080)          & (0.018)         & (0.075)        & (0.020)      
\end{tabular}
}
\end{table}

\begin{table}[p]
\centering
\caption{Quantitative performance of node-wise regression based on the 5 methods 
when $\varepsilon = 0.25$ and the tuning parameter
is selected by StARS. Each value shows the mean (standard deviation) on 100 simulated data sets.}
\label{tableS5}
\scalebox{0.8}{
\begin{tabular}{cc|cc|cc|cc|cc}
& & \multicolumn{2}{c|}{Chain} & \multicolumn{2}{c|}{Hub} & \multicolumn{2}{c|}{Scale-free} & \multicolumn{2}{c}{Random} \\ \hline
& & TPR         & FPR         & TPR        & FPR        & TPR            & FPR           & TPR          & FPR         \\ \hline
Asym. & $\gamma=0.3$ & 0.978       & 0.059       & 0.799      & 0.071      & 0.614          & 0.057         & 0.776        & 0.062       \\
           &       & (0.013)       & (0.024)       & (0.042)      & (0.020)      & (0.065)          & (0.019)         & (0.060)        & (0.021)       \\
           & $\gamma=0.5$ & 0.972       & 0.046       & 0.757      & 0.044      & 0.530          & 0.041         & 0.680        & 0.042       \\
           &       & (0.019)       & (0.016)       & (0.046)      & (0.019)      & (0.090)          & (0.019)         & (0.075)        & (0.017)       \\
           & Kend  & 0.391       & 0.031       & 0.241      & 0.034      & 0.136          & 0.031         & 0.188        & 0.036       \\
           &       & (0.087)       & (0.014)       & (0.071)      & (0.014)      & (0.051)          & (0.014)         & (0.066)        & (0.016)       \\
           & Grank & 0.367       & 0.030       & 0.228      & 0.031      & 0.132          & 0.029         & 0.174        & 0.032       \\
           &       & (0.088)       & (0.013)       & (0.071)      & (0.013)      & (0.051)          & (0.013)         & (0.065)        & (0.015)       \\
           & $Q_n$    & 0.881       & 0.062       & 0.642      & 0.063      & 0.429          & 0.058         & 0.547        & 0.059       \\
           &       & (0.043)       & (0.022)       & (0.063)      & (0.023)      & (0.069)          & (0.019)         & (0.069)        & (0.020)       \\ \hline
Sym.  & $\gamma=0.3$ & 0.977       & 0.063       & 0.806      & 0.075      & 0.622          & 0.059         & 0.782        & 0.066       \\
           &       & (0.016)       & (0.027)       & (0.038)      & (0.019)      & (0.062)          & (0.021)         & (0.047)        & (0.020)       \\
           & $\gamma=0.5$ & 0.975       & 0.047       & 0.765      & 0.047      & 0.526          & 0.041         & 0.678        & 0.040       \\
           &       & (0.018)       & (0.016)       & (0.053)      & (0.019)      & (0.084)          & (0.016)         & (0.067)        & (0.017)       \\
           & Kend  & 0.068       & 0.134       & 0.031      & 0.144      & 0.023          & 0.137         & 0.016        & 0.132       \\
           &       & (0.050)       & (0.019)       & (0.023)      & (0.017)      & (0.019)          & (0.020)         & (0.016)        & (0.020)       \\
           & Grank & 0.057       & 0.132       & 0.026      & 0.140      & 0.024          & 0.137         & 0.014        & 0.130       \\
           &       & (0.044)       & (0.020)       & (0.023)      & (0.019)      & (0.020)          & (0.020)         & (0.014)        & (0.019)       \\
           & $Q_n$    & 0.600       & 0.137       & 0.221      & 0.135      & 0.137          & 0.132         & 0.183        & 0.132       \\
           &       & (0.116)       & (0.018)       & (0.065)      & (0.013)      & (0.066)          & (0.015)         & (0.094)        & (0.019)      
\end{tabular}
}
\end{table}

\begin{table}[p]
\centering
\caption{Quantitative performance of CLIME based on the 5 methods 
when $\varepsilon = 0.05$ and the tuning parameter is selected by 2-fold cross validation. 
Each value shows the mean (standard deviation) on 100 simulated data sets.}
\label{tableS6}
\scalebox{0.7}{
\begin{tabular}{cc|ccc|ccc|ccc|ccc}
& & \multicolumn{3}{c|}{Chain} & \multicolumn{3}{c|}{Hub} & \multicolumn{3}{c|}{Scale-free} & \multicolumn{3}{c}{Random} \\ \hline
& & MSE     & TPR    & FPR    & MSE    & TPR    & FPR   & MSE      & TPR      & FPR      & MSE     & TPR     & FPR    \\ \hline
Asym. & $\gamma=0.3$ & 0.067   & 0.999  & 0.098  & 0.072  & 0.994  & 0.100 & 0.058    & 0.910    & 0.071    & 0.048   & 0.948   & 0.099  \\
      &       & (0.007)   & (0.005)  & (0.046)  & (0.006)  & (0.011)  & (0.039) & (0.006)    & (0.053)    & (0.039)    & (0.005)   & (0.027)   & (0.045)  \\
      & $\gamma=0.5$ & 0.105   & 0.995  & 0.034  & 0.115  & 0.976  & 0.045 & 0.079    & 0.789    & 0.032    & 0.064   & 0.840   & 0.037  \\
      &       & (0.006)   & (0.011)  & (0.023)  & (0.005)  & (0.022)  & (0.025) & (0.005)    & (0.098)    & (0.020)    & (0.005)   & (0.086)   & (0.024)  \\
      & Kend  & 0.105   & 0.997  & 0.067  & 0.112  & 0.974  & 0.073 & 0.084    & 0.797    & 0.047    & 0.068   & 0.872   & 0.062  \\
      &       & (0.007)   & (0.008)  & (0.033)  & (0.008)  & (0.026)  & (0.036) & (0.007)    & (0.109)    & (0.033)    & (0.005)   & (0.075)   & (0.036)  \\
      & Grank & 0.110   & 0.994  & 0.052  & 0.114  & 0.971  & 0.068 & 0.086    & 0.748    & 0.037    & 0.070   & 0.855   & 0.051  \\
      &       & (0.008)   & (0.012)  & (0.030)  & (0.008)  & (0.031)  & (0.033) & (0.008)    & (0.143)    & (0.030)    & (0.005)   & (0.070)   & (0.031)  \\
      & $Q_n$    & 0.129   & 0.999  & 0.057  & 0.135  & 0.992  & 0.086 & 0.095    & 0.865    & 0.044    & 0.077   & 0.922   & 0.058  \\
      &       & (0.003)   & (0.005)  & (0.029)  & (0.004)  & (0.013)  & (0.036) & (0.003)    & (0.070)    & (0.027)    & (0.003)   & (0.053)   & (0.031)  \\ \hline
Sym.  & $\gamma=0.3$ & 0.067   & 1.000  & 0.098  & 0.071  & 0.995  & 0.102 & 0.057    & 0.923    & 0.082    & 0.047   & 0.950   & 0.102  \\
      &       & (0.007)   & (0.005)  & (0.046)  & (0.006)  & (0.011)  & (0.039) & (0.006)    & (0.053)    & (0.039)    & (0.005)   & (0.027)   & (0.045)  \\
      & $\gamma=0.5$ & 0.104   & 0.996  & 0.038  & 0.114  & 0.980  & 0.050 & 0.080    & 0.773    & 0.030    & 0.063   & 0.866   & 0.044  \\
      &       & (0.006)   & (0.011)  & (0.023)  & (0.005)  & (0.022)  & (0.025) & (0.005)    & (0.098)    & (0.020)    & (0.005)   & (0.086)   & (0.024)  \\
      & Kend  & 0.107   & 0.995  & 0.074  & 0.114  & 0.964  & 0.086 & 0.085    & 0.756    & 0.045    & 0.070   & 0.831   & 0.059  \\
      &       & (0.007)   & (0.008)  & (0.033)  & (0.008)  & (0.026)  & (0.036) & (0.007)    & (0.109)    & (0.033)    & (0.005)   & (0.075)   & (0.036)  \\
      & Grank & 0.111   & 0.992  & 0.055  & 0.118  & 0.950  & 0.074 & 0.086    & 0.747    & 0.043    & 0.071   & 0.814   & 0.052  \\
      &       & (0.008)   & (0.012)  & (0.030)  & (0.008)  & (0.031)  & (0.033) & (0.008)    & (0.143)    & (0.030)    & (0.005)   & (0.070)   & (0.031)  \\
      & $Q_n$    & 0.130   & 0.999  & 0.057  & 0.137  & 0.989  & 0.078 & 0.096    & 0.842    & 0.046    & 0.078   & 0.915   & 0.067  \\
      &       & (0.003)   & (0.005)  & (0.029)  & (0.004)  & (0.013)  & (0.036) & (0.003)    & (0.070)    & (0.027)    & (0.003)   & (0.053)   & (0.031) 
\end{tabular}
}
\end{table}

\begin{table}[p]
\centering
\caption{Quantitative performance of CLIME based on the 5 methods 
when $\varepsilon = 0.15$ and the tuning parameter is selected by 2-fold cross validation. 
Each value shows the mean (standard deviation) on 100 simulated data sets.}
\label{tableS7}
\scalebox{0.7}{
\begin{tabular}{cc|ccc|ccc|ccc|ccc}
& & \multicolumn{3}{c|}{Chain} & \multicolumn{3}{c|}{Hub} & \multicolumn{3}{c|}{Scale-free} & \multicolumn{3}{c}{Random} \\ \hline
      &       & MSE     & TPR    & FPR    & MSE    & TPR    & FPR   & MSE      & TPR      & FPR      & MSE     & TPR     & FPR    \\ \hline
Asym. & $\gamma=0.3$ & 0.082   & 0.994  & 0.080  & 0.090  & 0.974  & 0.087 & 0.065    & 0.849    & 0.072    & 0.055   & 0.871   & 0.086  \\
      &       & (0.008)   & (0.011)  & (0.032)  & (0.009)  & (0.027)  & (0.035) & (0.006)    & (0.069)    & (0.035)    & (0.004)   & (0.045)   & (0.034)  \\
      & $\gamma=0.5$ & 0.104   & 0.991  & 0.054  & 0.113  & 0.946  & 0.065 & 0.080    & 0.731    & 0.043    & 0.065   & 0.780   & 0.047  \\
      &       & (0.007)   & (0.014)  & (0.029)  & (0.006)  & (0.032)  & (0.032) & (0.005)    & (0.113)    & (0.028)    & (0.004)   & (0.084)   & (0.027)  \\
      & Kend  & 0.164   & 0.770  & 0.030  & 0.166  & 0.623  & 0.033 & 0.122    & 0.302    & 0.019    & 0.100   & 0.339   & 0.022  \\
      &       & (0.005)   & (0.132)  & (0.022)  & (0.005)  & (0.165)  & (0.029) & (0.004)    & (0.149)    & (0.021)    & (0.003)   & (0.148)   & (0.018)  \\
      & Grank & 0.166   & 0.687  & 0.022  & 0.168  & 0.538  & 0.023 & 0.123    & 0.260    & 0.014    & 0.101   & 0.303   & 0.017  \\
      &       & (0.005)   & (0.179)  & (0.020)  & (0.005)  & (0.180)  & (0.023) & (0.004)    & (0.153)    & (0.016)    & (0.003)   & (0.132)   & (0.014)  \\
      & $Q_n$    & 0.162   & 0.989  & 0.075  & 0.162  & 0.929  & 0.076 & 0.122    & 0.756    & 0.059    & 0.100   & 0.817   & 0.074  \\
      &       & (0.006)   & (0.015)  & (0.034)  & (0.006)  & (0.042)  & (0.039) & (0.004)    & (0.096)    & (0.032)    & (0.003)   & (0.069)   & (0.034)  \\ \hline
Sym.  & $\gamma=0.3$ & 0.081   & 0.996  & 0.094  & 0.089  & 0.975  & 0.087 & 0.064    & 0.840    & 0.069    & 0.055   & 0.875   & 0.083  \\
      &       & (0.008)   & (0.011)  & (0.032)  & (0.009)  & (0.027)  & (0.035) & (0.006)    & (0.069)    & (0.035)    & (0.004)   & (0.045)   & (0.034)  \\
      & $\gamma=0.5$ & 0.105   & 0.991  & 0.053  & 0.114  & 0.950  & 0.067 & 0.080    & 0.715    & 0.038    & 0.065   & 0.777   & 0.049  \\
      &       & (0.007)   & (0.014)  & (0.029)  & (0.006)  & (0.032)  & (0.032) & (0.005)    & (0.113)    & (0.028)    & (0.004)   & (0.084)   & (0.027)  \\
      & Kend  & 0.158   & 0.669  & 0.124  & 0.160  & 0.502  & 0.136 & 0.115    & 0.241    & 0.096    & 0.095   & 0.268   & 0.109  \\
      &       & (0.005)   & (0.132)  & (0.022)  & (0.005)  & (0.165)  & (0.029) & (0.004)    & (0.149)    & (0.021)    & (0.003)   & (0.148)   & (0.018)  \\
      & Grank & 0.160   & 0.605  & 0.109  & 0.162  & 0.442  & 0.118 & 0.117    & 0.197    & 0.081    & 0.097   & 0.204   & 0.081  \\
      &       & (0.005)   & (0.179)  & (0.020)  & (0.005)  & (0.180)  & (0.023) & (0.004)    & (0.153)    & (0.016)    & (0.003)   & (0.132)   & (0.014)  \\
      & $Q_n$    & 0.168   & 0.949  & 0.140  & 0.171  & 0.822  & 0.151 & 0.128    & 0.596    & 0.115    & 0.105   & 0.627   & 0.137  \\
      &       & (0.006)   & (0.015)  & (0.034)  & (0.006)  & (0.042)  & (0.039) & (0.004)    & (0.096)    & (0.032)   & (0.003)  & (0.069)   & (0.034) 
\end{tabular}
}
\end{table}

\begin{table}[p]
\centering
\caption{Quantitative performance of CLIME based on the 5 methods 
when $\varepsilon = 0.25$ and the tuning parameter is selected by 2-fold cross validation. 
Each value shows the mean (standard deviation) on 100 simulated data sets.}
\label{tableS8}
\scalebox{0.7}{
\begin{tabular}{cc|ccc|ccc|ccc|ccc}
& & \multicolumn{3}{c|}{Chain} & \multicolumn{3}{c|}{Hub} & \multicolumn{3}{c|}{Scale-free} & \multicolumn{3}{c}{Random} \\ \hline
      &       & MSE     & TPR    & FPR    & MSE    & TPR    & FPR   & MSE      & TPR      & FPR      & MSE     & TPR     & FPR    \\ \hline
Asym. & $\gamma=0.3$ & 0.099   & 0.955  & 0.079  & 0.104  & 0.931  & 0.086 & 0.079    & 0.732    & 0.079    & 0.072   & 0.683   & 0.129  \\
      &       & (0.014)   & (0.149)  & (0.066)  & (0.013)  & (0.101)  & (0.076) & (0.012)    & (0.120)    & (0.115)    & (0.012)   & (0.251)   & (0.195)  \\
      & $\gamma=0.5$ & 0.111   & 0.951  & 0.048  & 0.119  & 0.885  & 0.059 & 0.084    & 0.599    & 0.037    & 0.070   & 0.627   & 0.041  \\
      &       & (0.008)   & (0.043)  & (0.029)  & (0.008)  & (0.069)  & (0.030) & (0.005)    & (0.105)    & (0.023)    & (0.005)   & (0.123)   & (0.026)  \\
      & Kend  & 0.197   & 0.154  & 0.009  & 0.195  & 0.107  & 0.009 & 0.148    & 0.049    & 0.006    & 0.122   & 0.048   & 0.006  \\
      &       & (0.003)   & (0.132)  & (0.014)  & (0.003)  & (0.102)  & (0.015) & (0.002)    & (0.059)    & (0.009)    & (0.002)   & (0.064)   & (0.009)  \\
      & Grank & 0.198   & 0.101  & 0.004  & 0.196  & 0.094  & 0.007 & 0.149    & 0.036    & 0.004    & 0.123   & 0.032   & 0.003  \\
      &       & (0.002)   & (0.103)  & (0.007)  & (0.003)  & (0.103)  & (0.013) & (0.002)    & (0.052)    & (0.008)    & (0.002)   & (0.043)   & (0.006)  \\
      & $Q_n$    & 0.192   & 0.878  & 0.100  & 0.191  & 0.768  & 0.097 & 0.149    & 0.519    & 0.074    & 0.121   & 0.544   & 0.089  \\
      &       & (0.004)   & (0.061)  & (0.048)  & (0.004)  & (0.075)  & (0.048) & (0.003)    & (0.103)    & (0.037)    & (0.003)   & (0.100)   & (0.045)  \\ \hline
Sym.  & $\gamma=0.3$ & 0.096   & 0.979  & 0.075  & 0.103  & 0.934  & 0.077 & 0.077    & 0.724    & 0.059    & 0.066   & 0.734   & 0.057  \\
      &       & (0.014)   & (0.149)  & (0.066)  & (0.013)  & (0.101)  & (0.076) & (0.012)    & (0.120)    & (0.115)    & (0.012)   & (0.251)   & (0.195)  \\
      & $\gamma=0.5$ & 0.112   & 0.958  & 0.048  & 0.119  & 0.886  & 0.054 & 0.084    & 0.581    & 0.034    & 0.070   & 0.615   & 0.040  \\
      &       & (0.008)   & (0.043)  & (0.029)  & (0.008)  & (0.069)  & (0.030) & (0.005)    & (0.105)    & (0.023)    & (0.005)   & (0.123)   & (0.026)  \\
      & Kend  & 0.171   & 0.592  & 0.319  & 0.171  & 0.436  & 0.337 & 0.124    & 0.308    & 0.301    & 0.102   & 0.336   & 0.316  \\
      &       & (0.003)   & (0.132)  & (0.014)  & (0.003)  & (0.102)  & (0.015) & (0.002)    & (0.059)    & (0.009)    & (0.002)   & (0.064)   & (0.009)  \\
      & Grank & 0.171   & 0.562  & 0.312  & 0.171  & 0.397  & 0.324 & 0.124    & 0.303    & 0.302    & 0.103   & 0.301   & 0.302  \\
      &       & (0.002)   & (0.103)  & (0.007)  & (0.003)  & (0.103)  & (0.013) & (0.002)    & (0.052)    & (0.008)    & (0.002)   & (0.043)   & (0.006)  \\
      & $Q_n$    & 0.203   & 0.697  & 0.172  & 0.203  & 0.313  & 0.144 & 0.156    & 0.301    & 0.157    & 0.131   & 0.138   & 0.096  \\
      &       & (0.004)   & (0.061)  & (0.048)  & (0.004)  & (0.075)  & (0.048) & (0.003)    & (0.103)    & (0.037)    & (0.003)   & (0.100)   & (0.045) 
\end{tabular}
}
\end{table}